\documentclass[final,1p]{elsarticle}
\usepackage{float}
\usepackage{ifpdf}
\usepackage{graphicx}
\usepackage{amssymb}
\usepackage{amsmath}

\journal{Communications in Nonlinear Science and Numerical Simulation}
\begin{document}

\begin{frontmatter} 

\title{Generalized Darboux transformation and N-th order rogue wave solution of a general coupled nonlinear Schr\"{o}dinger equations}
\author{N. Vishnu Priya}
\author{\corref{cor1}M. Senthilvelan}
\cortext[cor1]{Corresponding address: Centre for Nonlinear Dynamics, School of Physics,  Bharathidasan University, Tiruchirappalli - 620 024, Tamil Nadu, India}
\ead{velan@cnld.bdu.ac.in}
\address{Centre for Nonlinear Dynamics, School of Physics,  Bharathidasan University, \\Tiruchirappalli - 620 024, Tamil Nadu, India}





\begin{abstract}
We construct a generalized Darboux transformation (GDT) of a general coupled nonlinear Schr\"{o}dinger (GCNLS) system.  Using GDT method we derive a recursive formula and present determinant representations for N-th order rogue wave solution of this system.  Using these representations we derive first, second and third order rogue wave solutions with certain free parameters.  By varying these free parameters we demonstrate the formation of triplet, triangle and hexagonal patterns of rogue waves. 
\end{abstract}

\begin{keyword}
Coupled nonlinear Schr\"{o}dinger system \sep Generalized Darboux transformation \sep Rogue waves \sep Lax pair.

\MSC{37K40 \sep 35Q51 \sep 35Q55}
\end{keyword}

\end{frontmatter}

\section{Introduction}
\label{intro}
In recent years the study of rogue waves (RWs) got impetus due to their phenomenal properties and their use in potential applications, say for example supercontinuum generation in photonic crystal fibers and Bose-Einstein condensates \cite{{Pelin},{Akhmediev1}}.  RW is a localized object in both space and time  and appears from nowhere and disappears without a trace \cite{Akhmediev2}.  A wave is classified under this category when its wave height (distance from trough to crest) reaches a value which is at least twice the significant wave height \cite{{Pelin},{Akhmediev1},{Akhmediev2}}.  Even though it was first observed in arbitrary depth of ocean, the phenomenon is now shown to appear in diverse areas of physics including nonlinear optical fibers \cite{Solli}, BEC \cite{BEC}, super fluid He \cite{He}, capillary waves \cite{Capillary}, multi-component plasmas \cite{Plasma} and so on.  The most common mathematical description of RWs is based on certain rational solutions of the nonlinear Schr\"{o}dinger (NLS) equation, namely $i\psi_t+\psi_{xx}+2|\psi|^2\psi=0$, where $\psi$ represents amplitude of the wave and subscripts denote partial differentiation with respect to that variable.  Certain kinds of exact solutions of NLS equation have been considered to describe possible mechanism for the formation of RWs such as Peregrine soliton \cite{peregrine}, time periodic breather or Ma soliton (MS) \cite{{Ma},{Akhmediev3}} and space periodic breather or Akhmediev breather (AB) \cite{Akhmediev4}.  Subsequently attempts have been made to construct RW solutions through different methods for the NLS equation and its variants \cite{{Kalla},{Zhai},{Ling},{Degasperis},{Ohta},{Akhmediev5},{Akhmediev6},{Akhmediev7},{Akhmediev8},{Akhmediev9},{Akhmediev10},{Akhmediev11}}.  In this paper, we construct the N-th order RW solution of a general two coupled nonlinear Schr\"{o}dinger (GCNLS) system \cite{Wang},
\begin{subequations}
\begin{align}
ip_t+p_{xx}+2(a\vert p\vert ^{2}+c\vert q\vert ^{2}+bpq^{*}+b^{*}qp^{*})p &= 0, \\
iq_t+q_{xx}+2(a\vert p\vert ^{2}+c\vert q\vert ^{2}+bpq^{*}+b^{*}qp^{*})q &= 0,
\end{align}
\label{pct1}
\end{subequations}
where $p$ and $q$ are slowly varying pulse envelopes and $a$ and $c$ are real constants.  Here $b$ is a complex constant and $*$ denotes complex conjugation. The constants $a$ and $c$ describe the self phase modulation and cross phase modulation effects whereas the complex constant $b$ and $b^*$ describe the four wave mixing effects.  When $a=c$ and $b=0$, Eq. (\ref{pct1}) reduces to the well known Manakov system \cite{Manakov}.  When $a=-c$ and $b=0$ it reduces to the mixed coupled nonlinear Schr\"{o}dinger equation \cite{viji}.  
\par The Lax pair or eigenvalue problem of (\ref{pct1}) reads as
\begin{subequations}
\begin{align}
\Psi_x&=U\Psi  =(\Lambda J+P)\Psi,\\
\Psi_t&=V\Psi =(\Lambda^2V_0+\Lambda V_1+V_2)\Psi,
\label{pct2}
\end{align}
\label{pc2}
\end{subequations}
where $\Psi = (\psi(x,t), \phi(x,t), \varphi(x,t))^{T}$ is the vector eigenfunction and the superscript $T$ denotes the transpose of the matrix.  The block matrices $J$, $P$, $V_0$, $V_1$, $V_2$ and $\Lambda$ are given by\\
\begin{align}
J&= \begin{pmatrix}i&0&0\\0&i&0\\0&0&-i\end{pmatrix},\; P=\begin{pmatrix}0&0&p\\0&0&q\\r_1&r_2&0\end{pmatrix}, V_0 =-2i\begin{pmatrix}1&0&0\\0&1&0\\0&0&-1\end{pmatrix},\notag \\
V_1&=-2\begin{pmatrix}0&0&p\\0&0&q\\r_1&r_2&0\end{pmatrix},\;\; V_2 =\begin{pmatrix}-ipr_1&-ipr_2&ip_x\\-iqr_1&-iqr_2&iq_x\\-ir_{1x}&-ir_{2x}&ipr_1+iqr_2\end{pmatrix}
\label{pct3}
\end{align}
and $\Lambda=\text{diag}(\lambda,\lambda,\lambda)$ where $r_1=-(ap^*+bq^*)$, $r_2=-(b^*p^*+cq^*)$ and $\lambda$ is the isospectral parameter.  Eq. (\ref{pct1}) can be obtained from the zero curvature condition $U_t-V_x+[U,V]=0$, where the square  bracket denotes the usual commutator.  

\par Recently Wang et al \cite{Wang} have obtained N-bright soliton solution for the system (\ref{pct1}) through Riemann-Hilbert method and studied the collision dynamics between two bright solitons.  Later L\"{u} and Penj have examined the Painlev$\acute{e}$ integrability of this model and shown that it passes the Weiss-Tabor-Carnewell (WTC) algorithm \cite{Penj}.  The authors have also derived particular solutions through Painlev$\acute{e}$-B\"{a}cklund transformation.  Further, we have constructed dark-dark soliton, general breather (GB), Akhmediev breather (AB), Ma soliton (MS) and first order RW solutions of (\ref{pct1}) using Hirota bilinearization method. We have captured the dark-dark soliton in the defocusing regime and identified the other solutions, namely GB, AB, MS and RW in the focusing regime and also analyzed the characteristics of the above profiles with respect to the four-wave mixing parameter.  In addition to the above we have considered RW solution as the starting point and derived AB, MS and GB in the reverse direction.  In a follow-up work, we have constructed explicit higher order RW solutions of (\ref{pct1}) using modified Darboux transformation method (DT).  

\par Very recently efforts have also been made to construct N-th order RW solution of certain nonlinear evolution equations, see for example Refs. \cite{{liming1},{Porsezian},{zhaqilao1},{zhaqilao2},{liming2},{liming3},{liming4},{hqzhang}}.  Motivated by this contemporary development, in this paper, we intend to construct the N-th order RW solution of this model.  Since it is difficult to construct explicit N-th order RW solution using modified DT, we consider the generalized Darboux transformation method (GDT) and derive the N-th order RW solution of this system.  In the conventional N-fold DT one has N-distinct eigenvalues.  However, it has been shown that the higher order RW solutions do contain only one critical eigenvalue $\lambda_0$.  As our aim is to construct higher order RW solutions through DT we need to restrict all the eigenvalues $\lambda_i$, $i=1,2,...,N$, such that $\lambda_i\rightarrow\lambda_1=\lambda_0$.  It has been demonstrated that in the GDT, N distinct eigenvalues can be restricted to a single critical eigenvalue through a suitable limit process.  Applying this limit process successively we can construct the recursive formula for N-th order RW solution.  Applying the same limit process on the determinant representation of N-fold DT we can get the determinant representation of N-th order RW solution.  In this paper, we present both the recursive formula and determinant representation of N-th order RW solution.  We also analyze the structure of these RWs in detail with certain free parameters.  We present the explicit expression of first and second order RW solutions of (\ref{pct1}).  Since the explicit expression of third order RW solution is very lengthy we only give the third iterated GDT solution formula of it.  We derive the second and third order RW solutions with two and four free parameters respectively.  In addition to the above, we analyze the RW solutions based on these free parameters and obtain certain interesting structures.  For example, in the case of second order RW solution we get triplet structure and in the case of third order RW solution we observe triangular and hexagonal structures respectively.

The plan of the paper is as follows.  In Sec. \ref{DT}, we construct first, second and third iteration of DT for (\ref{pct1}) and present the N-th iteration of DT.  In Sec. (\ref{GDT}), we discuss GDT of Eq. (\ref{pct1}) in detail and present both recursive formula and determinant expressions of N-th order RW solution through GDT.  In Sec. (\ref{RW}), we derive the explicit form of first, second  and third order RW solutions.  A detailed description of first, second and third order RW solutions based on the free parameters is also included in this section.  Finally, in Sec. (\ref{conclusion}), we present our conclusions.
\section{Darboux transformation for GCNLS system}
\label{DT}
\subsection{First iteration}
A Darboux transformation (DT) is a special gauge transformation,
\begin{eqnarray}            
\Psi[1]=T[1]\Psi = \Psi\Lambda-S[1]\Psi,
\label{pct4}
\end{eqnarray}
where $\Psi$ and $\Psi[1]$ are old and new eigenfunctions of (\ref{pc2}), $T[1]$ is the Darboux matrix and $S[1]$ is a non-singular $3\times 3$ matrix.  The DT (\ref{pct4}) transforms the original Lax pair (\ref{pc2}) into a new Lax pair,
\begin{eqnarray}  
\Psi[1]_x=U[1]\Psi[1]=(\Lambda J +P[1])\Psi[1],\nonumber\\
\Psi[1]_t=V[1]\Psi[1]=(\Lambda^2V_0[1]+\Lambda V_1[1]+V_2[1])\Psi[1],
\label{pct5}
\end{eqnarray}
in which the matrices $P[1]$, $V_0[1]$, $V_1[1]$ and $V_2[1]$ assume the same forms as that of $P$, $V_0$, $V_1$ and $V_2$ except that the potentials $p$ and $q$ have now acquired new expressions, namely $p[1]$ and $q[1]$ in $U[1]$ and $V[1]$.  Substituting the transformation (\ref{pct4}) into (\ref{pc2}) and comparing the resultant expressions with (\ref{pct5}), we find
\begin{eqnarray}            
U[1]=(T[1]_x+T[1]U)T[1]^{-1},\;\; V[1]=(T[1]_t+T[1]V)T[1]^{-1}.
\label{pct6}
\end{eqnarray}
Plugging the expressions $U[1]$, $V[1]$, $U$, $V$ and $T[1]$ in Eq. (\ref{pct6}) and equating the coefficients of various powers of $\Lambda$ on both sides we get the following relations between old and new potentials, namely
\begin{subequations}
\begin{align}
V_0[1]&= V_0, \label{pct10} \\
V_1[1]&= V_1+[V_0,S[1]],\label{pct11} \\
V_2[1]&= V_2+[V_1,S[1]]+[V_0,S[1]]S[1],\label{pct12} \\
P[1] & = P+[J,S[1]],\label{pct13} \\
S[1]_x & = [P,S[1]]+[J,S[1]]S[1], \label{pct14} \\
S[1]_t & = [V_2,S[1]]+[V_1,S[1]]S[1]+[V_0,S[1]]S[1]^2.
\label{pct15}
\end{align}
\label{pc15}
\end{subequations}
The eigenvalue problem given in (\ref{pc2}) remains invariant under the transformation (\ref{pct4}) provided $S[1]$ satisfies all the Eqs. (\ref{pct10})-(\ref{pct15}).

We assume a general form for the matrix $S[1]$, namely
\begin{align}
 S[1]=\begin{pmatrix}S_{11}&S_{12}&S_{13}\\S_{21}&S_{22}&S_{23}\\S_{31}&S_{32}&S_{33}\end{pmatrix}.
\label{pctS}
\end{align}
Substituting the assumed form of $S[1]$ in Eq. (\ref{pct13}) and equating the matrix elements on both sides, we find
\begin{eqnarray}
p[1]=p+2iS_{13}, \;\; q[1]=q+2iS_{23}.
\label{p1q1}
\end{eqnarray}
\par To determine the exact form of $S[1]$ we consider $S[1]$ to be \cite{Matveev},
\begin{align}
S[1]=H_1\Lambda_1 H_1^{-1},
\label{pct18}
\end{align}
where
\begin{eqnarray}
H_1=\left(\begin{array}{ccc}\psi_1&\varphi_1^*&0\\\phi_1&0&\varphi_1^*
\\\varphi_1
&-\psi_1^*&-\phi_1^*\end{array}\right), \quad  \Lambda_1=\left(\begin{array}{ccc}\lambda_1&0&0\\0&\lambda_1^*&0\\0&0&\lambda_1^*\end{array}\right).
\label{pct8}
\end{eqnarray}
In Eq. (\ref{pct8}) $\Psi_1=\left(\psi_1, \phi_1, \varphi_1\right)^T$ is the column solution of Lax pair equations (\ref{pc2}) for the initial potential ($p$, $q$) at $\lambda=\lambda_1$.  Then it follows from the orthogonality condition that $\left(\varphi_1^*,0,-\psi_1^*\right)^T$ and $\left(0, \varphi_1^*, -\phi_1^*\right)^T$ are also the solutions of (\ref{pc2}) at  $\lambda=\lambda_1^*$.  
\par The first iterated DT is given by $\Psi[1]=T[1]\Psi=\Psi\Lambda-S[1]\Psi$ (vide Eq.(\ref{pct4})).  If $H_1$ is the solution of $\Psi[1]$ at $\Lambda=\Lambda_1$ then it should satisfy \cite{Matveev}
\begin{align}
\Psi_1[1]=T[1]H_1=0.
\label{pct21}
\end{align}
In other words
\begin{align}
H_1\Lambda_1-S[1] H_1=0 \Rightarrow S[1] H_1=H_1\Lambda_1.
\label{pct22}
\end{align}
Expressing Eq. (\ref{pct22}) in matrix form, we have
\begin{align}
\begin{pmatrix}
S_{11}&S_{12}&S_{13}\\
S_{21}&S_{22}&S_{23}\\
S_{31}&S_{32}&S_{33}
\end{pmatrix}
\times &
\begin{pmatrix}
\psi_1&\varphi_1^*&0\\
\phi_1&0&\varphi_1^*\\
\varphi_1&-\psi_1^*&-\phi_1^*
\end{pmatrix} \notag \\
 &=\begin{pmatrix}
\lambda_1\psi_1&\lambda_1^*\varphi_1^*&0\\
\lambda_1\phi_1&0&\lambda_1^*\varphi_1^*\\
\lambda_1\varphi_1&-\lambda_1^*\psi_1^*&-\lambda_1^*\phi_1^*
\end{pmatrix} .
\label{pct23}
\end{align}
\par Using Cramer's rule we can determine the exact expression of $S_{13}$ and $S_{23}$ which in turn reads
\begin{eqnarray}
S_{13}=\frac{\left|\begin{array}{ccc}\psi_1&\varphi_1^*&0\\\phi_1&0&\varphi_1^*
\\\lambda_1\psi_1&\lambda_1^*\varphi_1^*&0\end{array}\right|}{\left|\begin{array}{ccc}\psi_1&\varphi_1^*&0\\\phi_1&0&\varphi_1^* 
\\ \varphi_1&-\psi_1^*&-\phi_1^*\end{array}\right|},\;\;\;
S_{23}=\frac{\left|\begin{array}{ccc}\psi_1&\varphi_1^*&0\\\phi_1&0&\varphi_1^*
\\\lambda_1\phi_1&0&\lambda_1^*\varphi_1^*\end{array}\right|}{\left|\begin{array}{ccc}\psi_1&\varphi_1^*&0\\\phi_1&0&\varphi_1^*
\\\varphi_1
&-\psi_1^*&-\phi_1^*\end{array}\right|}.
\end{eqnarray}
Evaluating the above determinants, we find
\begin{subequations}
\begin{align}
S_{13}&= \displaystyle{\frac{(\lambda_1-\lambda_1^*)\psi_1\varphi_1^*}{|\psi_1|^2+|\phi_1|^2+|\varphi_1|^2}}, \\
S_{23}&= \displaystyle{\frac{(\lambda_1-\lambda_1^*)\phi_1\varphi_1^*}{|\psi_1|^2+|\phi_1|^2+|\varphi_1|^2}}.
\end{align}
\label{pct24}
\end{subequations}
From (\ref{pct24}) it is evident that to determine $S_{13}$ and $S_{23}$ one should know the explicit expressions of $\psi_1$, $\phi_1$ and $\varphi_1$ which are the solutions of the eigenvalue problem (\ref{pc2}).  They can be determined by solving the following six coupled linear first order partial differential equations, namely
\begin{align}
\psi_{1x} = & i\lambda_1\psi_1+p\varphi_1,\notag\\
\phi_{1x} = & i\lambda_1\phi_1+q\varphi_1,\notag\\
\varphi_{1x} = & r_1\psi_1+r_2\phi_1-i\lambda \varphi_1,\notag\\
\psi_{1t} = & (-2i\lambda_1^2-ipr_1)\psi_1-ipr_2\phi_1+(ip_x-2p\lambda_1)\varphi_1,\notag\\
\phi_{1t} = & (-2i\lambda_1^2-iqr_2)\phi_1-iqr_1\psi_1+(iq_x-2q\lambda_1)\varphi_1,\notag\\
\varphi_{1t} = & (-ir_{1x}-2\lambda_1r_1)\psi_1+(-ir_{2x}-2\lambda_1r_2)\phi_1\notag\\
& +(ipr_1+iqr_2+2i\lambda_1^2)\varphi_1.
\label{pct256}
\end{align}
\par Solving the system of equations given in (\ref{pct256}) with appropriate seed solution $p$ and $q$, one can obtain the explicit expressions of $\psi_1$, $\phi_1$ and $\varphi_1$.  With the known expressions of $\psi_1$, $\phi_1$ and $\varphi_1$ the matrix elements $S_{13}$ and $S_{23}$ can now be fixed.  Plugging the latter into (\ref{p1q1}), we obtain the solution for the Eq. (\ref{pct1}) in the form
\begin{eqnarray}
p[1]&=&p+2i\frac{(\lambda_1-\lambda_1^*)\psi_1\varphi_1^*}{|\psi_1|^2+|\phi_1|^2+|\varphi_1|^2},\nonumber\\
q[1]&=&q+2i\frac{(\lambda_1-\lambda_1^*)\phi_1\varphi_1^*}{|\psi_1|^2+|\phi_1|^2+|\varphi_1|^2}.
\label{pct9}
\end{eqnarray}
Eq. (\ref{pct9}) can also be written in a more compact determinant form, that is
\begin{eqnarray}
p[1]=p+2i \frac{\left|\begin{array}{ccc}\psi_1&\varphi_1^*&0\\\phi_1&0&\varphi_1^*
\\\lambda_1\psi_1&\lambda_1^*\varphi_1^*&0\end{array}\right|}{\left|\begin{array}{ccc}\psi_1&\varphi_1^*&0\\\phi_1&0&\varphi_1^*
\\\varphi_1
&-\psi_1^*&-\phi_1^*\end{array}\right|},\;\;\;
q[1]=q+2i \frac{\left|\begin{array}{ccc}\psi_1&\varphi_1^*&0\\\phi_1&0&\varphi_1^*
\\\lambda_1\phi_1&0&\lambda_1^*\varphi_1^*\end{array}\right|}{\left|\begin{array}{ccc}\psi_1&\varphi_1^*&0\\\phi_1&0&\varphi_1^*
\\\varphi_1
&-\psi_1^*&-\phi_1^*\end{array}\right|}.
\label{fidet}
\end{eqnarray}
Through the formula (\ref{pct9}) or (\ref{fidet}) one can generate a class of solutions including solitary wave solution, breather and RW solution for the Eq. (\ref{pct1}).  
\subsection{Second iteration}
The second iterated DT reads,
\begin{eqnarray}            
\Psi[2]=T[2]\Psi[1] = \Psi[1]\Lambda-S[2]\Psi[1],
\label{pc21}
\end{eqnarray}
where $\Psi[2]$ and $\Psi[1]$ are the second and first iterated eigenfunctions respectively and $T[2]$ is the second iterated DT matrix.  The DT (\ref{pc21}) transforms the first iterated Lax pair (\ref{pct5}) into the second iterated Lax pair of the same type
\begin{eqnarray}  
\Psi[2]_x=U[2]\Psi[2]=(\Lambda J +P[2])\Psi[2],\nonumber\\
\Psi[2]_t=V[2]\Psi[2]=(\Lambda^2 V_0[2]+\Lambda V_1[2]+V_2[2])\Psi[2],
\label{pc22}
\end{eqnarray}
where the matrices $P[2]$, $V_0[2]$, $V_1[2]$ and $V_2[2]$ have the same forms as that of $P[1]$, $V_0[1]$, $V_1[1]$ and $V_2[1]$ except that the potentials $p[1]$ and $q[1]$ have now acquired new expressions $p[2]$ and $q[2]$ in the matrices $U[2]$ and $V[2]$. Substituting the transformation (\ref{pc21}) into (\ref{pc22}) and rearranging the resultant equations, we get 
\begin{eqnarray}            
U[2]=(T[2]_x+T[2]U[1])T[2]^{-1},\;\; V[2]=(T[2]_t+T[2]V[1])T[2]^{-1}.
\label{pc23}
\end{eqnarray}

\par Substituting the matrix expressions of $U[2]$, $V[2]$, $U[1]$, $V[1]$ and $T[2]$ in Eq. (\ref{pc23}) and equating the coefficients of various powers of $\Lambda$ on both sides we get the following relations between the first iterated and second iterated potentials, namely
\begin{subequations}
\begin{align}
V_0[2]&= V_0[1], \label{pct241} \\
V_1[2]&= V_1[1]+[V_0[1],S[2]],\label{pct242} \\
V_2[2]&= V_2[1]+[V_1[1],S[2]]+[V_0[1],S[2]]S[2],\label{pct243} \\
P[2] & = P[1]+[J,S[2]],\label{pct244} \\
S[2]_x & = [P[1],S[2]]+[J,S[2]]S[2], \label{pct245} \\
S[2]_t & = [V_2[1],S[2]]+[V_1[1],S[2]]S[2]+[V_0[1],S[2]]S[2]^2.
\label{pct246}
\end{align}
\label{pc24}
\end{subequations}
The eigenvalue problem given in (\ref{pc22}) remains invariant under the transformation (\ref{pc21}), provided $S[2]$ satisfies all the Eqs. (\ref{pct241})-(\ref{pct246}). 

\par We consider the matrix $S[2]$ is of the form
\begin{align}
S[2]=\begin{pmatrix}S[2]_{11}&S[2]_{12}&S[2]_{13}\\S[2]_{21}&S[2]_{22}&S[2]_{23}\\S[2]_{31}&S[2]_{32}&S[2]_{33}\end{pmatrix}
\label{pc25}
\end{align}
so that upon substituting (\ref{pc25}) in Eq. (\ref{pct244}) we can find the following two expressions which link the new and old potentials respectively, that is
\begin{eqnarray}
p[2]=p[1]+2iS[2]_{13}, \;\; q[2]=q[1]+2iS[2]_{23}.
\label{pc26}
\end{eqnarray}
\par We determine the elements in the matrix $S[2]$ by assuming the latter to be
\begin{align}
S[2]=H_2[1]\Lambda_2 H_2[1]^{-1},
\label{pc27}
\end{align}
where
\begin{eqnarray}
H_2[1]=\left(\begin{array}{ccc}\psi_2[1]&\varphi_2[1]^*&0\\\phi_2[1]&0&\varphi_2[1]^*
\\\varphi_2[1]
&-\psi_2[1]^*&-\phi_2[1]^*\end{array}\right), \quad  \Lambda_2=\left(\begin{array}{ccc}\lambda_2&0&0\\0&\lambda_2^*&0\\0&0&\lambda_2^*\end{array}\right).
\label{pc28}
\end{eqnarray}
In Eq. (\ref{pc28}) $\Psi_2[1]=\left(\psi_2[1], \phi_2[1], \varphi_2[1]\right)^T$ is the column solution of the Lax pair (\ref{pc22}) at $\lambda=\lambda_2$.  It follows from the orthogonality condition that $\left(\varphi_2[1]^*,0,-\psi_2[1]^*\right)^T$ and $\left(0, \varphi_2[1]^*, -\phi_2[1]^*\right)^T$ are also the solutions of (\ref{pc22}) at  $\lambda=\lambda_2^*$.  If $H_2[1]$ is the solution of $\Psi[2]$ at $\Lambda=\Lambda_2$ then it should satisfy
\begin{align}
\Psi[2]=T[2]H_2[1]=0.
\label{pc29}
\end{align}
In other words
\begin{align}
H_2[1]\Lambda_2-S[2] H_2[1]=0 \Rightarrow S[2] H_2[1]=H_2[1]\Lambda_2.
\label{pc29}
\end{align}
The explicit matrix form of  Eq. (\ref{pc29}) is given by 
\begin{align}
\begin{pmatrix}
S[2]_{11}&S[2]_{12}&S[2]_{13}\\
S[2]_{21}&S[2]_{22}&S[2]_{23}\\
S[2]_{31}&S[2]_{32}&S[2]_{33}
\end{pmatrix}
\times &
\begin{pmatrix}
\psi_2[1]&\varphi_2[1]^*&0\\
\phi_2[1]&0&\varphi_2[1]^*\\
\varphi_2[1]&-\psi_2[1]^*&-\phi_2[1]^*
\end{pmatrix} \notag \\
 &=\begin{pmatrix}
\lambda_2\psi_2[1]&\lambda_2^*\varphi_2[1]^*&0\\
\lambda_2\phi_2[1]&0&\lambda_2^*\varphi_2[1]^*\\
\lambda_2\varphi_2[1]&-\lambda_2^*\psi_2[1]^*&-\lambda_2^*\phi_2[1]^*
\end{pmatrix}.
\label{pc30}
\end{align}
\par Since we need to know the exact expression of only two matrix elements, namely $S[2]_{13}$, $S[2]_{23}$ (vide Eq.(\ref{pc26})), we determine them from the above equation with the help of Cramer's rule.  Our result shows that
\begin{eqnarray}
S[2]_{13}&=&\frac{(\lambda_2-\lambda_2^*)\psi_2[1]\varphi_2^*[1]}{|\psi_2[1]|^2+|\phi_2[1]|^2+|\varphi_2[1]|^2}, \nonumber\\
S[2]_{23}&=&\frac{(\lambda_2-\lambda_2^*)\phi_2[1]\varphi_2^*[1]}{|\psi_2[1]|^2+|\phi_2[1]|^2+|\varphi_2[1]|^2}.
\label{pc31}
\end{eqnarray}

From (\ref{pc31}) it is evident that to determine $S[2]_{13}$ and $S[2]_{23}$ one should know the explicit expressions of $\psi_2[1]$, $\phi_2[1]$ and $\varphi_2[1]$ which are nothing but the solutions of the Lax pair equations,
\begin{align}
\psi_2[1]_{x} = & i\lambda_2\psi_2[1]+p[1]\varphi_2[1],\notag\\
\phi_2[1]_{x} = & i\lambda_2\phi_2[1]+q[1]\varphi_2[1],\notag\\
\varphi_2[1]_{x} = & r_1[1]\psi_2[1]+r_2[1]\phi_2[1]-i\lambda_2 \varphi_2[1],\notag\\
\psi_2[1]_{t} = & (-2i\lambda_2^2-ip[1]r_1[1])\psi_2[1]-ip[1]r_2[1]\phi_2[1]+(ip[1]_x-2p[1]\lambda_2)\varphi_2[1],\notag\\
\phi_2[1]_{t} = & (-2i\lambda_2^2-iq[1]r_2[1])\phi_2[1]-iq[1]r_1[1]\psi_2[1]+(iq[1]_x-2q[1]\lambda_2)\varphi_2[1],\notag\\
\varphi_2[1]_{t} = & (-ir_1[1]_{x}-2\lambda_2r_1[1])\psi_2[1]+(-ir_2[1]_x-2\lambda_2r_2[1])\phi_2[1]\notag\\
& +(ip[1]r_1[1]+iq[1]r_2[1]+2i\lambda_2^2)\varphi_2[1],
\label{pc256}
\end{align}
with $r_1[1]=-(ap[1]^*+bq[1]^*)$, $r_2=-(b^*p[1]^*+cq[1]^*)$ where $p[1]$ and $q[1]$ are the first iterated potentials.

\par Solving Eq. (\ref{pc256}) consistently with first iterated potentials $p[1]$ and $q[1]$ we can obtain the exact forms of $\psi_2[1]$, $\phi_2[1]$ and $\varphi_2[1]$.  Substituting them in (\ref{pc31}) we can get the matrix elements $S[2]_{13}$ and $S[2]_{23}$.  Plugging the latter into (\ref{pc26}) we arrive at the second iterated solution for the Eq. (\ref{pct1}) in the form
\begin{eqnarray}
p[2]&=&p[1]+2i\frac{(\lambda_2-\lambda_2^*)\psi_2[1]\varphi_2^*[1]}{|\psi_2[1]|^2+|\phi_2[1]|^2+|\varphi_2[1]|^2},\nonumber\\
q[2]&=&q[1]+2i\frac{(\lambda_2-\lambda_2^*)\phi_2[1]\varphi_2^*[1]}{|\psi_2[1]|^2+|\phi_2[1]|^2+|\varphi_2[1]|^2}.
\label{pc32}
\end{eqnarray} 
Using (\ref{pc32}) one can generate a class of solutions including 2-soliton solutions, second order breathers and so on for the Eq. (\ref{pct1}).  Eq. (\ref{pc32}) can also be written in a more compact form, that is
\begin{eqnarray}
p[2]=p+2i\displaystyle{\frac{|N_1[2]|}{|D[2]|}},  \;\; q[2]=q+2i\displaystyle{\frac{|N_2[2]|}{|D[2]|}},
\label{pd331}
\end{eqnarray} 
where the matrices $N_1[2]$, $N_2[2]$ and $D[2]$ are given by
\begin{eqnarray}
N_1[2]=&\left(\begin{array}{cccccc}\lambda_1\psi_1&\lambda_2\psi_2&\lambda_1^*\varphi_1^*&\lambda_2^*\varphi_2^*&0&0\\\psi_1&\psi_2&\varphi_1^*&\varphi_2^*&0&0\\\lambda_1\phi_1&\lambda_2\phi_2&0&0&\lambda_1^*\varphi_1^*&\lambda_2^*\varphi_2^*\\\phi_1&\phi_2&0&0&\varphi_1^*&\varphi_2^*\\\lambda_1^2\psi_1&\lambda_2^2\psi_2&\lambda_1^{*2}\varphi_1^*&\lambda_2^{*2}\varphi_2^*&0&0\\\varphi_1&\varphi_2&-\psi_1^*&-\psi_2^*&-\phi_1^*&-\phi_2^*\end{array}\right),
\nonumber
\end{eqnarray}

\begin{eqnarray}
N_2[2]=&\left(\begin{array}{cccccc}\lambda_1\psi_1&\lambda_2\psi_2&\lambda_1^*\varphi_1^*&\lambda_2^*\varphi_2^*&0&0\\\psi_1&\psi_2&\varphi_1^*&\varphi_2^*&0&0\\\lambda_1\phi_1&\lambda_2\phi_2&0&0&\lambda_1^*\varphi_1^*&\lambda_2^*\varphi_2^*\\\phi_1&\phi_2&0&0&\varphi_1^*&\varphi_2^*\\\lambda_1^2\phi_1&\lambda_2^2\phi_2&0&0&\lambda_1^{*2}\varphi_1^*&\lambda_2^{*2}\varphi_2^*\\\varphi_1&\varphi_2&-\psi_1^*&-\psi_2^*&-\phi_1^*&-\phi_2^*\end{array}\right),
\nonumber
\end{eqnarray}
and 
\begin{eqnarray}
D[2]=&\left(\begin{array}{cccccc}\lambda_1\psi_1&\lambda_2\psi_2&\lambda_1^*\varphi_1^*&\lambda_2^*\varphi_2^*&0&0\\\psi_1&\psi_2&\varphi_1^*&\varphi_2^*&0&0\\\lambda_1\phi_1&\lambda_2\phi_2&0&0&\lambda_1^*\varphi_1^*&\lambda_2^*\varphi_2^*\\\phi_1&\phi_2&0&0&\varphi_1^*&\varphi_2^*\\\lambda_1\varphi_1&\lambda_2\varphi_2&-\lambda_1^*\psi_1^*&-\lambda_2^*\psi_2^*&-\lambda_1^*\phi_1^*&-\lambda_2^*\phi_2^*\\\varphi_1&\varphi_2&-\psi_1^*&-\psi_2^*&-\phi_1^*&-\phi_2^*\end{array}\right).
\label{pd3311}
\end{eqnarray}
In the above determinant representations $\left(\psi_2,\phi_2,\varphi_2\right)^T$ is the column solution of Lax pair equations (\ref{pc2}) at $\lambda=\lambda_2$ so that  $\left(\varphi_2^*,0,-\psi_2^*\right)^T$ and $\left(0, \varphi_2^*, -\phi_2^*\right)^T$ are also the solutions of (\ref{pc2}) at  $\lambda=\lambda_2^*$.  
\subsection{Third Iteration}
Since we plan to construct third order RW solution also, in the following, we recall the essential expressions to derive it.  The third iterated DT matrix,
\begin{eqnarray}
T[3]=\Lambda-H_3[2]\Lambda[3] H_3[2]^{-1},\label{pcT32}
\end{eqnarray}
acts on the eigenfunction $\Psi[2]$ yields
\begin{eqnarray}
\Psi[3]=T[3]\Psi[2].
\end{eqnarray}
This transformation changes the second iterated solution $p[2]$ and $q[2]$ into new solution $p[3]$ and $q[3]$ as
\begin{eqnarray}
p[3]&=&p[2]+2i\frac{(\lambda_3-\lambda_3^*)\psi_3[2]\varphi_3^*[2]}{|\psi_3[2]|^2+|\phi_3[2]|^2+|\varphi_3[2]|^2},\nonumber\\
q[3]&=&q[2]+2i\frac{(\lambda_3-\lambda_3^*)\phi_3[2]\varphi_3^*[2]}{|\psi_3[2]|^2+|\phi_3[2]|^2+|\varphi_3[2]|^2},
\label{pcdt33}
\end{eqnarray} 
where $(\psi_3[2],\phi_3[2],\varphi_3[2])^T$ is the column solution of the Lax pair equation at $\lambda=\lambda_3$ with $p[2]$ and $q[2]$ as the seed solution.  From the formula (\ref{pcdt33}) we can derive three soliton solution and third order breather solution using suitable seed solution. 
Using the ideas given in the previous subsection we can rewrite Eq. (\ref{pcdt33}) in a more compact form,
\begin{eqnarray}
p[3]=p+2i\displaystyle{\frac{|N_1[3]|}{|D[3]|}},  \;\; q[3]=q+2i\displaystyle{\frac{|N_2[3]|}{|D[3]|}},
\label{pd332}
\end{eqnarray} 
where the matrices $N_1[3]$, $N_2[3]$ and $D[3]$ are given by
\begin{eqnarray}
N_1[3]=&&\left(\begin{array}{ccccccccc}\lambda_1^2\psi_1&\lambda_2^2\psi_2&\lambda_3^2\psi_3&\lambda_1^{*2}\varphi_1^*&\lambda_2^{2*}\varphi_2^*&\lambda_3^{2*}\varphi_3^*&0&0&0\\\lambda_1\psi_1&\lambda_2\psi_2&\lambda_3\psi_3&\lambda_1^{*}\varphi_1^*&\lambda_2^{*}\varphi_2^*&\lambda_3^{*}\varphi_3^*&0&0&0\\\psi_1&\psi_2&\psi_3&\varphi_1^*&\varphi_2^*&\varphi_3^*&0&0&0\\\lambda_1^2\phi_1&\lambda_2^2\phi_2&\lambda_3^2\phi_3&0&0&0&\lambda_1^{*2}\varphi_1^*&\lambda_2^{*2}\varphi_2^*&\lambda_3^{*2}\varphi_3^*\\\lambda_1\phi_1&\lambda_2\phi_2&\lambda_3\phi_3&0&0&0&\lambda_1^{*}\varphi_1^*&\lambda_2^{*}\varphi_2^*&\lambda_3^{*}\varphi_3^*\\\phi_1&\phi_2&\phi_3&0&0&0&\varphi_1^*&\varphi_2^*&\varphi_3^*\\\lambda_1^3\psi_1&\lambda_2^3\psi_2&\lambda_3^3\psi_3&\lambda_1^{*3}\varphi_1^*&\lambda_2^{*3}\varphi_2^*&\lambda_3^{*3}\varphi_3^*&0&0&0\\\lambda_1\varphi_1&\lambda_2\varphi_2&\lambda_3\varphi_3&-\lambda_1^{*}\psi_1^*&-\lambda_2^{*}\psi_2^*&-\lambda_3^{*}\psi_3^*&-\lambda_1^{*}\phi_1^*&-\lambda_2^{*}\phi_2^*&-\lambda_3^{*}\phi_3^*\\\varphi_1&\varphi_2&\varphi_3&-\psi_1^*&-\psi_2^*&-\psi_3^*&-\phi_1^*&-\phi_2^*&-\phi_3^*\end{array}\right),\nonumber
\end{eqnarray}
\begin{eqnarray}
N_2[3]=&\left(\begin{array}{ccccccccc}\lambda_1^2\psi_1&\lambda_2^2\psi_2&\lambda_3^2\psi_3&\lambda_1^{*2}\varphi_1^*&\lambda_2^{2*}\varphi_2^*&\lambda_3^{2*}\varphi_3^*&0&0&0\\\lambda_1\psi_1&\lambda_2\psi_2&\lambda_3\psi_3&\lambda_1^{*}\varphi_1^*&\lambda_2^{*}\varphi_2^*&\lambda_3^{*}\varphi_3^*&0&0&0\\\psi_1&\psi_2&\psi_3&\varphi_1^*&\varphi_2^*&\varphi_3^*&0&0&0\\\lambda_1^2\phi_1&\lambda_2^2\phi_2&\lambda_3^2\phi_3&0&0&0&\lambda_1^{*2}\varphi_1^*&\lambda_2^{*2}\varphi_2^*&\lambda_3^{*2}\varphi_3^*\\\lambda_1\phi_1&\lambda_2\phi_2&\lambda_3\phi_3&0&0&0&\lambda_1^{*}\varphi_1^*&\lambda_2^{*}\varphi_2^*&\lambda_3^{*}\varphi_3^*\\\phi_1&\phi_2&\phi_3&0&0&0&\varphi_1^*&\varphi_2^*&\varphi_3^*\\\lambda_1^3\phi_1&\lambda_2^3\phi_2&\lambda_3^3\phi_3&0&0&0&\lambda_1^{*3}\varphi_1^*&\lambda_2^{*3}\varphi_2^*&\lambda_3^{*3}\varphi_3^*\\\lambda_1\varphi_1&\lambda_2\varphi_2&\lambda_3\varphi_3&-\lambda_1^{*}\psi_1^*&-\lambda_2^{*}\psi_2^*&-\lambda_3^{*}\psi_3^*&-\lambda_1^{*}\phi_1^*&-\lambda_2^{*}\phi_2^*&-\lambda_3^{*}\phi_3^*\\\varphi_1&\varphi_2&\varphi_3&-\psi_1^*&-\psi_2^*&-\psi_3^*&-\phi_1^*&-\phi_2^*&-\phi_3^*\end{array}\right)\nonumber
\end{eqnarray}
and
\begin{eqnarray}
D[3]=&\left(\begin{array}{ccccccccc}\lambda_1^2\psi_1&\lambda_2^2\psi_2&\lambda_3^2\psi_3&\lambda_1^{*2}\varphi_1^*&\lambda_2^{2*}\varphi_2^*&\lambda_3^{2*}\varphi_3^*&0&0&0\\\lambda_1\psi_1&\lambda_2\psi_2&\lambda_3\psi_3&\lambda_1^{*}\varphi_1^*&\lambda_2^{*}\varphi_2^*&\lambda_3^{*}\varphi_3^*&0&0&0\\\psi_1&\psi_2&\psi_3&\varphi_1^*&\varphi_2^*&\varphi_3^*&0&0&0\\\lambda_1^2\phi_1&\lambda_2^2\phi_2&\lambda_3^2\phi_3&0&0&0&\lambda_1^{*2}\varphi_1^*&\lambda_2^{*2}\varphi_2^*&\lambda_3^{*2}\varphi_3^*\\\lambda_1\phi_1&\lambda_2\phi_2&\lambda_3\phi_3&0&0&0&\lambda_1^{*}\varphi_1^*&\lambda_2^{*}\varphi_2^*&\lambda_3^{*}\varphi_3^*\\\phi_1&\phi_2&\phi_3&0&0&0&\varphi_1^*&\varphi_2^*&\varphi_3^*\\\lambda_1^2\varphi_1&\lambda_2^2\varphi_2&\lambda_3^2\varphi_3&-\lambda_1^{*2}\psi_1^*&-\lambda_2^{*2}\psi_2^*&-\lambda_3^{*2}\psi_3^*&-\lambda_1^{*2}\phi_1^*&-\lambda_2^{*2}\phi_2^*&-\lambda_3^{*2}\psi_3^*\\\lambda_1\varphi_1&\lambda_2\varphi_2&\lambda_3\varphi_3&-\lambda_1^{*}\psi_1^*&-\lambda_2^{*}\psi_2^*&-\lambda_3^{*}\psi_3^*&-\lambda_1^{*}\phi_1^*&-\lambda_2^{*}\phi_2^*&-\lambda_3^{*}\phi_3^*\\\varphi_1&\varphi_2&\varphi_3&-\psi_1^*&-\psi_2^*&-\psi_3^*&-\phi_1^*&-\phi_2^*&-\phi_3^*\end{array}\right).\nonumber\\
\end{eqnarray}
In the above determinant representations $\left(\psi_3,\phi_3,\varphi_3\right)^T$ is the column solution of Lax pair equations (\ref{pc2}) at $\lambda=\lambda_3$ and $\left(\varphi_3^*,0,-\psi_3^*\right)^T$ and $\left(0, \varphi_3^*, -\phi_3^*\right)^T$ are solutions of (\ref{pc2}) at  $\lambda=\lambda_3^*$.  
\subsection{N-th Iteration}
If $N$ distinct basic solutions $\Psi_k=(\psi_k,\phi_k,\varphi_k)^T$, $k=1,2,...,N$, of the Lax pair (\ref{pc2}) are given at $\lambda=\lambda_k$, $k=1,2,...,N$, then the N-fold DT can be iterated successively.  The $N$-th iterated solution turns out to be
\begin{eqnarray}
p[N]=p[N-1]+2i\frac{(\lambda_N-\lambda_N^*)\psi_N[N-1]\varphi_N^*[N-1]}{|\psi_N[N-1]|^2+|\phi_N[N-1]|^2+|\varphi_N[N-1]|^2},\nonumber
\end{eqnarray}
\begin{eqnarray}
q[N]=q[N-1]+2i\frac{(\lambda_N-\lambda_N^*)\phi_N[N-1]\varphi_N^*[N-1]}{|\psi_N[N-1]|^2+|\phi_N[N-1]|^2+|\varphi_N[N-1]|^2},
\label{pctnth}
\end{eqnarray} 
where $\Psi[N-1]=T[N-1]\Psi[N-2]=T[N-1]T[N-2]...T[1]T[0]\Psi$ and \\
$T[N]=\Lambda_{N}-H_N[N-1]\Lambda[N] H_N[N-1]^{-1}$ with 
\begin{eqnarray}
\Lambda[N]&=&\left(\begin{array}{ccc}\lambda_N&0&0\\0&\lambda_N^*&0\\0&0&\lambda_N^*\end{array}\right),\nonumber
\end{eqnarray}
\begin{eqnarray}
H_N[N-1]=\left(\begin{array}{ccc}\psi_N[N-1]&\varphi_N^*[N-1]&0\\\phi_N[N-1]&0&\varphi_N^*[N-1]
\\\varphi_N[N-1]&-\psi_N^*[N-1]&-\phi_N^*[N-1]\end{array}\right).
\end{eqnarray}
By solving the Lax pair equations with suitable seed solution and substituting the obtained forms in the solution formula (\ref{pctnth}) we can construct a class of solutions including N-soliton solutions, N-breather solutions and so on.
\par The determinant forms of the $N$-th iterated solutions are given by,
\begin{eqnarray}
p[N]=p[0]+2i\frac{\Delta_2}{\Delta_1},\qquad q[N]=q[0]+2i\frac{\Delta_3}{\Delta_1},\label{nth}
\end{eqnarray}
\begin{eqnarray}
\Delta_1 = \scriptsize\left|\begin{array}{ccccccccc}
\lambda_1^{N-1}\psi_1&\cdots&{\lambda}_N^{N-1} \psi_N & \lambda_1^{*(N-1)}\varphi_1^* &\cdots &\lambda_N^{*(N-1)}\varphi_N^*& 0&\cdots&0 \\
\cdots&\cdots&\cdots&\cdots&\cdots&\cdots&\cdots&\cdots&\cdots \\\psi_1 & \cdots &\psi_N & \varphi_1^* & \cdots &\varphi_N^*& 0 &\cdots&0 \\\lambda_1^{N-1}\phi_1&\cdots&\lambda_N^{N-1}\phi_N & 0&\cdots &0& \lambda_1^{*(N-1)}\varphi_1^* & \cdots& \lambda_N^{*(N-1)}\varphi_N^*\\\cdots&\cdots&\cdots&\cdots&\cdots&\cdots&\cdots&\cdots&\cdots \\\phi_1 & \cdots& \phi_N & 0&\cdots &0& \varphi_1^* & \cdots& \varphi_N^*\\
\lambda_1^{N-1}\varphi_1  &\cdots&\lambda_N^{N-1}\varphi_N & -\lambda_1^{*(N-1)}\psi_1^*& \cdots&-\lambda_N^{*(N-1)}\psi_N^*  & -\lambda_1^{*(N-1)} \phi_1^*  &\cdots&-\lambda_N^{*(N-1)}\phi_N^*  \\\cdots&\cdots&\cdots&\cdots&\cdots&\cdots&\cdots&\cdots&\cdots\\\varphi_1 & \cdots& \varphi_N & -\psi_1^* & \cdots& -\psi_N^*& -\phi_1^* & \cdots& -\phi_N^*
\end{array} \right|,\nonumber
\end{eqnarray}
\begin{eqnarray}
\Delta_2 = \scriptsize\left|\begin{array}{ccccccccc}
\lambda_1^{N-1}\psi_1&\cdots&{\lambda}_N^{N-1} \psi_N & \lambda_1^{*(N-1)}\varphi_1^* &\cdots &\lambda_N^{*(N-1)}\varphi_N^*& 0&\cdots&0 \\
\cdots&\cdots&\cdots&\cdots&\cdots&\cdots&\cdots&\cdots&\cdots \\\psi_1 & \cdots &\psi_N & \varphi_1^* & \cdots &\varphi_N^*& 0 &\cdots&0 \\\lambda_1^{N-1}\phi_1&\cdots&\lambda_N^{N-1}\phi_N & 0&\cdots &0& \lambda_1^{*(N-1)}\varphi_1^* & \cdots& \lambda_N^{*(N-1)}\varphi_N^*\\\cdots&\cdots&\cdots&\cdots&\cdots&\cdots&\cdots&\cdots&\cdots \\\phi_1 & \cdots& \phi_N & 0&\cdots &0& \varphi_1^* & \cdots& \varphi_N^*\\
\lambda_1^{N}\psi_1&\cdots&{\lambda}_N^{N} \psi_N & \lambda_1^{*(N)}\varphi_1^* &\cdots &\lambda_N^{*(N)}\varphi_N^*& 0&\cdots&0  \\\cdots&\cdots&\cdots&\cdots&\cdots&\cdots&\cdots&\cdots&\cdots\\\varphi_1 & \cdots& \varphi_N & -\psi_1^* & \cdots& -\psi_N^*& -\phi_1^* & \cdots& -\phi_N^*
\end{array} \right|,\nonumber
\end{eqnarray}
\begin{eqnarray}
\Delta_3 = \scriptsize\left|\begin{array}{ccccccccc}
\lambda_1^{N-1}\psi_1&\cdots&{\lambda}_N^{N-1} \psi_N & \lambda_1^{*(N-1)}\varphi_1^* &\cdots &\lambda_N^{*(N-1)}\varphi_N^*& 0&\cdots&0 \\
\cdots&\cdots&\cdots&\cdots&\cdots&\cdots&\cdots&\cdots&\cdots \\\psi_1 & \cdots &\psi_N & \varphi_1^* & \cdots &\varphi_N^*& 0 &\cdots&0 \\\lambda_1^{N-1}\phi_1&\cdots&\lambda_N^{N-1}\phi_N & 0&\cdots &0& \lambda_1^{*(N-1)}\varphi_1^* & \cdots& \lambda_N^{*(N-1)}\varphi_N^*\\\cdots&\cdots&\cdots&\cdots&\cdots&\cdots&\cdots&\cdots&\cdots \\\phi_1 & \cdots& \phi_N & 0&\cdots &0& \varphi_1^* & \cdots& \varphi_N^*\\
\lambda_1^{N}\phi_1&\cdots&\lambda_N^{N}\phi_N & 0&\cdots &0& \lambda_1^{*(N)}\varphi_1^* & \cdots& \lambda_N^{*(N)}\varphi_N^* \\\cdots&\cdots&\cdots&\cdots&\cdots&\cdots&\cdots&\cdots&\cdots\\\varphi_1 & \cdots& \varphi_N & -\psi_1^* & \cdots& -\psi_N^*& -\phi_1^* & \cdots& -\phi_N^*
\end{array} \right|.\nonumber
\end{eqnarray}
Eq. (\ref{nth}) gives N-th iterated DT solution formula from which one can construct N-soliton solution and N-breather solution explicitly.
\section{Generalized Darboux transformation (GDT)}
\label{GDT}
We have shown that the N-th iterated DT contains N-eigenfunctions $\Psi_i$, $i=1,2,...,N,$ associated with N-distinct eigenvalues $\lambda_i$, $i=1,2,...,N$.  In the conventional DT the N-th iteration T[N] annihilates its generating eigenfunction.  In other words, the generating eigenfunction $\Psi_i$ cannot be used more than once when we carry out the iteration in the original DT scheme.  However, to obtain a higher order RW solution for a critical eigenvalue $\lambda_0$ we must apply repeated DTs.  As mentioned in \cite{Porsezian} this difficulty can be overcome by noting that the annihilated eigenfunctions can be regained if we take the limit $\lambda_i\to\lambda_1$ in the corresponding eigenvalues found in the DT.  Adopting the procedure developed in Ref. \cite{liming1} we generate higher order RW solutions which involve only one eigenvalue, namely $\lambda_0$.
\subsection{First iteration of GDT}
From the classical Darboux theory, we infer that $T[1]\Psi_1=0$ and so we cannot apply DT on $\Psi_1$ again.  Suppose $\Psi_2=\Psi_1(\lambda_1+\delta)$ is a special solution for the Lax pair (\ref{pc2}).  Then $\frac{\Psi_1(\lambda_1+\delta)}{\delta}$, where $\delta$ is a small parameter, is also a solution of (\ref{pc2}).  Expanding the eigenfunction $\Psi_2$ in Taylor series at $\lambda_1$, we get
\begin{eqnarray}
\Psi_2=\Psi_1(\lambda_1+\delta)=\Psi_1^{[0]}+\Psi_1^{[1]}\delta+\Psi_1^{[2]}\delta^2+...+\Psi_1^{[N]}\delta^N+...,
\label{g1}
\end{eqnarray}
where $\Psi_1^{[k]}=\frac{1}{k!}\frac{\partial^k}{\partial\lambda^k}\Psi_1(\lambda)|_{\lambda=\lambda_1}, \; k=0,1,2,\cdots$.  Since $\Psi_1^{[0]}$ is the solution of (\ref{pc2}) at $\lambda=\lambda_1$ with initial seed solutions $p$ and $q$ the first step GDT of (\ref{pct1}) turns out that
\begin{eqnarray}
p[1]&=&p+2i\frac{(\lambda_1-\lambda_1^*)\psi_1^{[0]}\varphi_1^{*[0]}}{|\psi_1^{[0]}|^2+|\phi_1^{[0]}|^2+|\varphi_1^{[0]}|^2},\nonumber\\
q[1]&=&q+2i\frac{(\lambda_1-\lambda_1^*)\phi_1^{[0]}\varphi_1^{*[0]}}{|\psi_1^{[0]}|^2+|\phi_1^{[0]}|^2+|\varphi_1^{[0]}|^2}.
\label{g2}
\end{eqnarray} 
As far as the first iterated solution is concerned one may note that there is no difference between the conventional DT and GDT.
\subsection{Second iteration}
Now we make the second iterated GDT through the limit process \cite{liming1}.  Doing so, we find
\begin{eqnarray}
{\lim_{\delta \rightarrow 0}}\frac{T[1]|_{\lambda=\lambda_1+\delta}\Psi_2}{\delta}&=&{\lim_{\delta \rightarrow 0}}\frac{[\delta+T[1]|_{\lambda=\lambda_1}]\Psi_1(\lambda_1+\delta)}{\delta}\nonumber\\
&=&\Psi_1^{[0]}+T[1]|_{\lambda=\lambda_1}\Psi_1^{[1]}=\Psi_1[1],
\label{g3}
\end{eqnarray}
where $\Psi_1^{[0]}$ and $\Psi_1^{[1]}$ are already determined through the expansion given in (\ref{g1}) and $\Psi_1[1]$ is the new iterated eigenfunction obtained through GDT which is a function of $\lambda_1$ only.  Since $T[1]$ is also known, the right hand side of (\ref{g3}) provides the exact form of $\Psi_1[1]=(\psi_1[1], \phi_1[1], \varphi_1[1])^T$.  Substituting them in (\ref{pc32}) we arrive at the second iterated GDT solution of (\ref{pct1}) in the limit $\lambda_2\rightarrow\lambda_1$, that is
\begin{eqnarray}
p[2]&=&p[1]+2i\frac{(\lambda_1-\lambda_1^*)\psi_1[1]\varphi_1^*[1]}{|\psi_1[1]|^2+|\phi_1[1]|^2+|\varphi_1[1]|^2},\nonumber\\ 
q[2]&=&q[1]+2i\frac{(\lambda_1-\lambda_1^*)\phi_1[1]\varphi_1^*[1]}{|\psi_1[1]|^2+|\phi_1[1]|^2+|\varphi_1[1]|^2}.
\label{2ndgdt}
\end{eqnarray} 

\par Performing the limit process on the determinants given in Eqs. (\ref{pd331}) and (\ref{pd3311}) we can get the determinant form of the solution (\ref{2ndgdt}).  By comparing the Eqs. (\ref{pc32}) and (\ref{2ndgdt}), we  observe that the eigenvalues and their corresponding eigenfunctions that appear in (\ref{pc32}) are replaced by the eigenvalue $\lambda_1$ and its associated eigenfunction. 
To express (\ref{2ndgdt}) in the determinant form we need to perform the same limit process directly on (\ref{pd331}).  Doing so, we find  
\begin{eqnarray}
p[2]=p+2i\displaystyle{\frac{|M_1[2]|}{|H[2]|}}|_{\lambda_2\rightarrow\lambda_1},  \;\; q[2]=q+2i\displaystyle{\frac{|M_2[2]|}{|H[2]|}}|_{\lambda_2\rightarrow\lambda_1},
\label{sd133}
\end{eqnarray} 
where
\begin{eqnarray}
H[2]|_{\lambda_2\rightarrow\lambda_1}=&\left(\begin{array}{cccccc}\lambda_1\psi_1&\psi_1[1,1]&\lambda_1^*\varphi_1^*&0&\varphi_1[1,1]^*&0\\\psi_1&\psi_1[0,1]&\varphi_1^*&0&\varphi_1[0,1]^*&0\\\lambda_1\phi_1&\phi_1[1,1]&0&-\lambda_1^*\varphi_1^*&0&\varphi_1[1,1]^*\\\phi_1&\phi_1[0,1]&0&\varphi_1^*&0&\varphi_1^*[0,1]\\\lambda_1\varphi_1&\varphi_1[1,1]&\lambda_1^*\psi_1^*&-\lambda_1^*\phi_1^*&-\psi_1^*[1,1]&-\phi_1^*[1,1]\\\varphi_1&\varphi_1[0,1]&-\psi_1^*&-\phi_1^*&-\psi_1^*[0,1]&-\phi_1[0,1]^*\end{array}\right),
\nonumber
\end{eqnarray}

\begin{eqnarray}
M_1[2]|_{\lambda_2\rightarrow\lambda_1}=&\left(\begin{array}{cccccc}\lambda_1\psi_1&\psi_1[1,1]&\lambda_!^*\varphi_1^*&0&\varphi_1[1,1]*&0\\\psi_1&\psi_1[0,1]&\varphi_1^*&0&\varphi_1[0,1]^*&0\\\lambda_1\phi_1&\phi_1[1,1]&0&-\lambda_1^*\varphi_1^*&0&\varphi_1[1,1]^*\\\phi_1&\phi_1[0,1]&0&\varphi_1^*&0&\varphi_1^*[0,1]\\\lambda_1^2\psi_1&\psi_1[2,1]&\lambda_1^{*2}\varphi_1^*&0&-\varphi_1^*[2,1]&0\\\varphi_1&\varphi_1[0,1]&-\psi_1^*&-\phi_1^*&-\psi_1^*[0,1]&-\phi_1[0,1]^*\end{array}\right).
\nonumber
\end{eqnarray}
\begin{eqnarray}
M_2[2]|_{\lambda_2\rightarrow\lambda_1}=&\left(\begin{array}{cccccc}\lambda_1\psi_1&\psi_1[1,1]&\lambda_1^*\varphi_1^*&0&\varphi_1[1,1]^*&0\\\psi_1&\psi_1[0,1]&\varphi_1^*&0&\varphi_1[0,1]^*&0\\\lambda_1\phi_1&\phi_1[1,1]&0&-\lambda_1^*\varphi_1^*&0&\varphi_1[1,1]^*\\\phi_1&\phi_1[0,1]&0&\varphi_1^*&0&\varphi_1^*[0,1]\\\lambda_1^2\phi_1&\phi_1[2,1]&0&\lambda_1^2\varphi_1^*&0&\varphi_1[2,1]^*\\\varphi_1&\varphi_1[0,1]&-\psi_1^*&\phi_1^*&-\psi_1^*[0,1]&-\phi_1[0,1]^*\end{array}\right).
\nonumber
\end{eqnarray}
with
\begin{eqnarray}
\psi_1[j,n]&=&\frac{1}{n!}\frac{\partial^{n}}{\partial \lambda_1^{n}}[(\lambda_1+\delta)^j\psi_1(\lambda_1+\delta)]|_{\delta=0},\nonumber\\
\phi_1[j,n]&=&\frac{1}{n!}\frac{\partial^{n}}{\partial \lambda_1^{n}}[(\lambda_1+\delta)^j\phi_1(\lambda_1+\delta)]|_{\delta=0},\nonumber\\
\varphi_1[j,n]&=&\frac{1}{n!}\frac{\partial^{n}}{\partial \lambda_1^{n}}[(\lambda_1+\delta)^j\varphi_1(\lambda_1+\delta)]|_{\delta=0},\;\; j,n = 0,1,2.
\label{sd133111}
\end{eqnarray}
Using the determinant representation (\ref{sd133}) and (\ref{sd133111}) we can straightforwardly derive the second order RW solution of GCNLS system (\ref{pct1}).
\subsection{Third iteration}
In this subsection we generate the third iteration through limit process.  The resultant action yields
\begin{eqnarray}
{\lim_{\delta \rightarrow 0}}\frac{[T_1[2]|_{\lambda_1+\delta}][\frac{\delta+T_1[1](\lambda_1)}{\delta}]\Psi_2}{\delta}
&=&\Psi_1^{[0]}+[T_1[1](\lambda_1)+T_1[2](\lambda_1)]\Psi_1^{[1]}\nonumber\\
&&+T_1[2](\lambda_1)T_1[1](\lambda_1)\Psi_1^{[2]} \nonumber\\
&=&\Psi_1[2],
\label{pctpsi12}
\end{eqnarray}
where $\Psi_1^{[0]}$, $\Psi_1^{[1]}$ and $\Psi_1^{[2]}$ are given in Eq. (\ref{g1}) and $\Psi_1[2]$ is the second iterated eigenfunction obtained through GDT which is a function of $\lambda_1$ only.  Since $T_1[1]$ and $T_1[2]$ are also known, the limit process (\ref{pctpsi12}) straightforwardly gives a nontrivial solution to the Lax pair (\ref{pc2}) with $p[2]$ and $q[2]$ as the seed solution with only one eigenvalue, that is $\lambda_1$.  The final form of third iterated GDT solution reads
\begin{eqnarray}
p[3]&=&p[2]+2i\frac{(\lambda_1-\lambda_1^*)\psi_1[2]\varphi_1^*[2]}{|\psi_1[2]|^2+|\phi_1[2]|^2+|\varphi_1[2]|^2},\nonumber\\ q[3]&=&q[2]+2i\frac{(\lambda_1-\lambda_1^*)\phi_1[2]\varphi_1^*[2]}{|\psi_1[2]|^2+|\phi_1[2]|^2+|\varphi_1[2]|^2},
\label{pct3rd}
\end{eqnarray} 
where
$\Psi_1[2]=\Psi_1^{[0]}+[T_1[1](\lambda_1)+T_1[2](\lambda_1)]\Psi_1^{[1]}+T_1[2](\lambda_1)T_1[1](\lambda_1)\Psi_1^{[2]}$ and $T_1[2]=\Lambda[1]-H_1[2]\Lambda[1] H_1[2]^{-1}$ with
\begin{eqnarray}
H_1[2]&=&\left(\begin{array}{ccc}\psi_1[2]&\varphi_1^*[2]&0\\\phi_1[2]&0&\varphi_1^*[2]\\\varphi_1[2]
&-\psi_1^*[2]&-\phi_1^*[2]\end{array}\right).
\end{eqnarray}
\par Applying the limit process on the determinants (\ref{pd332}) we can get the third iterated solution in the following form:
\begin{eqnarray}
p[3]=p+2i\displaystyle{\frac{|M_1[3]|}{|H[3]|}}|_{\lambda_2\rightarrow\lambda_1},  \;\; q[3]=q+2i\displaystyle{\frac{|M_2[3]|}{|H[3]|}}|_{\lambda_2\rightarrow\lambda_1},
\label{sd233}
\end{eqnarray}
where
\begin{eqnarray}
|M_1[3]|&=&\left(\begin{array}{ccc}A&B&C\\D&E&F\\G_1&H&J\end{array}\right),\;\;|M_2[3]|=\left(\begin{array}{ccc}A&B&C\\D&E&F\\G_2&H&J\end{array}\right)\nonumber\\|H[3]|&=&\left(\begin{array}{ccc}A&B&C\\D&E&F\\G_3&H_2&J_2\end{array}\right)
\end{eqnarray}
and $A,B,C,D,E,F,G,H,J$ are block matrices whose explicit forms are given by
\begin{eqnarray}
A&=&\left(\begin{array}{ccc}\lambda_1^2\psi_1&\psi_1[2,1]&\psi_1[2,2]\\\lambda_1\psi_1&\psi_1[1,1]&\psi_1[1,2]\\\psi_1&\psi_1[0,1]&\psi_1[0,2]\end{array}\right), \;B=\left(\begin{array}{ccc}\lambda_1^{*2}\varphi_1^*&0&\varphi_1[2,1]*\\\lambda_1^*\varphi_1^*&0&\varphi_[1,1]^*\\\varphi_1^*&0&\varphi_[0,1]^*\end{array}\right),\nonumber\\ 
C&=&\left(\begin{array}{ccc}0&\varphi_1[2,2]*&0\\0&\varphi_[1,2]^*&0\\0&\varphi_[0,2]^*&0\end{array}\right),\; D=\left(\begin{array}{ccc}\lambda_1^2\phi_1&\phi_1[2,1]&\phi_1[2,2]\\\lambda_1\phi_1&\phi_1[1,1]&\phi_1[1,2]\\\phi_1&\phi_1[0,1]&\phi_1[0,2]\end{array}\right),\nonumber\\
E&=&\left(\begin{array}{ccc}0&-\lambda_1^{2*}\varphi_1^*&0\\0&-\lambda^{*}\varphi_1^*&0\\0&-\varphi_1^*&0\end{array}\right),\; F=\left(\begin{array}{ccc}\varphi_1[2,1]^*&0&\varphi_1[2,2]^*\\\varphi_1[1,1]^*&0&\varphi_1[1,2]^*\\\varphi_1[0,1]^*&0&\varphi_1[0,2]^*\end{array}\right),\nonumber
\end{eqnarray}
\begin{eqnarray}
G_1&=&\left(\begin{array}{ccc}\lambda_1^{*3}\phi_1&\phi_1[3,1]&\phi_1[3,2]\\\lambda_1^*\varphi_1&\varphi_1[1,1]&\varphi_1[1,2]\\\varphi_1&\varphi_1[0,1]&\varphi_1[0,2]\end{array}\right), \;
H=\left(\begin{array}{ccc}0&\lambda_1^{*3}\varphi_1^*&0\\-\lambda_1\psi_1^*&-\lambda_1\phi_1^*&-\psi_1^*[1,1]\\-\psi_1^*&-\phi_1^*&-\psi_1[0,1]^*\end{array}\right),\nonumber\\
J&=&\left(\begin{array}{ccc}-\varphi_1^*[3,1]&0&\varphi_1[3,2]^*\\-\phi_1[1,1]^*&-\psi_1^*[1,2]&-\phi_1[1,2]^*\\-\phi_1[0,1]^*&-\psi_1^*[0,2]&-\phi_1[0,2]^*\end{array}\right),\; G_2=\left(\begin{array}{ccc}\lambda_1^{*3}\psi_1&\psi_1[3,1]&\psi_1[3,2]\\\lambda_1^*\varphi_1&\varphi_1[1,1]&\varphi_1[1,2]\\\varphi_1&\varphi_1[0,1]&\varphi_1[0,2]\end{array}\right),\nonumber\\
G_3&=&\left(\begin{array}{ccc}\lambda_1^{*2}\varphi_1&\varphi_1[2,1]&\varphi_1[2,2]\\\lambda_1^*\varphi_1&\varphi_1[1,1]&\varphi_1[1,2]\\\varphi_1&\varphi_1[0,1]&\varphi_1[0,2]\end{array}\right),\; 
H_2=\left(\begin{array}{ccc}-\lambda_1^{*2}\psi_1^*&\lambda_1^{*2}\phi_1^*&-\psi_1^*[2,1]\\-\lambda_1\psi_1^*&-\lambda_1\phi_1^*&-\psi_1^*[1,1]\\-\psi_1^*&-\phi_1^*&-\psi_1[0,1]^*\end{array}\right),\nonumber\\J_2&=&-\left(\begin{array}{ccc}\phi_1^*[2,1]&\psi_1[2,2]^*&\phi_1[2,1]^*\\\phi_1[1,1]^*&\psi_1^*[1,2]&\phi_1[1,2]^*\\\phi_1[0,1]^*&\psi_1^*[0,2]&\phi_1[0,2]^*\end{array}\right),
\end{eqnarray}
with
\begin{eqnarray}
\psi_1[j,n]&=&\frac{1}{n!}\frac{\partial^{n}}{\partial \lambda_1^{n}}[(\lambda_1+\delta)^j\psi_1(\lambda_1+\delta)]|_{\delta=0},\nonumber\\
\phi_1[j,n]&=&\frac{1}{n!}\frac{\partial^{n}}{\partial \lambda_1^{n}}[(\lambda_1+\delta)^j\phi_1(\lambda_1+\delta)]|_{\delta=0},\nonumber\\
\varphi_1[j,n]&=&\frac{1}{n!}\frac{\partial^{n}}{\partial \lambda_1^{n}}[(\lambda_1+\delta)^j\varphi_1(\lambda_1+\delta)]|_{\delta=0},\;\; j,n = 0,1,2,3.
\label{sd13311}
\end{eqnarray}
The determinant representation (\ref{sd233}) also provides third order RW solution of (\ref{pct1}).
\subsection{N-th iteration of GDT}
Continuing the above limit process and combining all the Darboux matrices, we can constitute the N-th step GDT which is of the form 
\begin{eqnarray}
p[N]=p[N-1]+2i\frac{(\lambda_1-\lambda_1^*)\psi_1[N-1]\varphi_1^*[N-1]}{|\psi_1[N-1]|^2+|\phi_1[N-1]|^2+|\varphi_1[N-1]|^2},\nonumber
\end{eqnarray}
\begin{eqnarray}
q[N]=q[N-1]+2i\frac{(\lambda_1-\lambda_1^*)\phi_1[N-1]\varphi_1^*[N-1]}{|\psi_1[N-1]|^2+|\phi_1[N-1]|^2+|\varphi_1[N-1]|^2},
\label{rec1}
\end{eqnarray} 
with
\begin{eqnarray}
\Psi_1[N-1]&=&\Psi_1^{[0]}+\left[\sum_{l=1}^{N-1}T_1[l](\lambda_1)\right]\Psi_1^{[1]}+\left[\sum_{l=1}^k\sum_{l<k}^{N-1}T_1[k](\lambda_1)T_1[l]\lambda_1\right]\Psi_1^{[2]}\nonumber\\&&\;\;+...+\left[T_1[N-1](\lambda_1)T_1[N-2](\lambda_1)...T_1[1](\lambda_1)\right]\Psi_1^{[n]} \nonumber 
\end{eqnarray}
and $T_1[k]=\Lambda[1]-H_1[k-1]\Lambda[1] H_1[k-1]^{-1}$.  The matrix $H_1[k-1]$ is defined by
\begin{eqnarray}
H_1[k-1]=\left(\begin{array}{ccc}\psi_1[k-1]&\varphi_1^*[k-1]&0\\\phi_1[k-1]&0&\varphi_1^*[k-1]
\\\varphi_1[k-1]
&-\psi_1^*[k-1]&-\phi_1^*[k-1]\end{array}\right).\label{rec2}
\end{eqnarray}
The expressions (\ref{rec1}) and (\ref{rec2}) turn out to be as N-th order RW solution of (\ref{pct1}). 
\par To present the determinant form of the N-th iterated GDT we perform the limit process on the determinants given in N-th iterated classical DT (\ref{nth}).  Assuming that N-distinct solutions $\Psi_i=(\psi_i,\phi_i,\varphi_i)^T,\; i=1,2,...,n$, are given for the Lax pair (\ref{pc2}) at $\lambda =\lambda _1,...,\lambda_i=\lambda_n$, and expanding the solutions in Taylor series in the following form
\begin{eqnarray}
(\lambda_i+\delta)^j\psi_i(\lambda_i+\delta)&=&\lambda_i^j\psi_i+\psi_i[j,1]\delta+\psi_i[j,2]\delta^2+...+\psi_i[j,m_i]\delta^{m_i}+...,\nonumber\\
(\lambda_i+\delta)^j\phi_i(\lambda_i+\delta)&=&\lambda_i^j\phi_i+\phi_i[j,1]\delta+\phi_i[j,2]\delta^2+...+\phi_i[j,m_i]\delta^{m_i}+...,\nonumber\\
(\lambda_i+\delta)^j\varphi_i(\lambda_i+\delta)&=&\lambda_i^j\varphi_i+\varphi_i[j,1]\delta+\varphi_i[j,2]\delta^2+...+\varphi_i[j,m_i]\delta^{m_i}+...,
\label{series}
\end{eqnarray}
where 
\begin{eqnarray}
\psi_1[j,m]&=&\frac{1}{m!}\frac{\partial^m}{\partial\lambda^m}[\lambda^j\Psi_i(\lambda)]|_{\lambda=\lambda_i},\nonumber\\
 \phi_1[j,m]&=&\frac{1}{m!}\frac{\partial^m}{\partial\lambda^m}[\lambda^j\Phi_i(\lambda)]|_{\lambda=\lambda_i},\nonumber\\
 \varphi_1[j,m]&=&\frac{1}{m!}\frac{\partial^m}{\partial\lambda^m}[\lambda^j\Phi_i(\lambda)]|_{\lambda=\lambda_i},\;\; j=0,1,2,...N, \; m=1,2,...N,
\nonumber
\end{eqnarray}
and applying the limit process on the determinants (\ref{nth}), we find
\begin{eqnarray}
p[N]=p+2i\frac{D_2}{D_1},\quad q[N]=q+2i\frac{D_3}{D_1}.
\label{gdtnth}
\end{eqnarray}
The determinants $D_1$, $D_2$ and $D_3$ are given by
\begin{eqnarray}
&&D_1 =\nonumber\\&&{\tiny\left|\begin{array}{ccccccccc}
 (i\sqrt{c/2})^{N-1}\psi_1& \cdots &\psi_1[N-1,N-1]& (i\sqrt{c/2})^{N-1} \varphi_1^*& \cdots& \varphi_1^*[N-1,N-1]& 0 &\cdots&0\\
\cdots&\cdots&\cdots&\cdots&\cdots&\cdots&\cdots&\cdots&\cdots\\\psi_1 & \cdots &\psi_1[0,N-1] & \varphi_1^* & \cdots& \varphi_1^*[0,N-1]& 0 &\cdots&0\\(i\sqrt{c/2})^{N-1}\phi_1&\cdots&\phi_1 [N-1,N-1]& 0& \cdots&0 &\varphi_1^* (i\sqrt{c/2})^{N-1}& \cdots&\varphi_1^*[N-1,N-1] \\\cdots&\cdots&\cdots&\cdots&\cdots&\cdots&\cdots&\cdots&\cdots\\\phi_1 & \cdots& \phi_1[0,N-1]& 0 &\cdots&0 & \varphi_1^* & \cdots& \varphi_1^*[0,N-1]\\
(i\sqrt{c/2})^{N-1} \varphi_1 & \cdots&\varphi_1[N-1,N-1] & -(i\sqrt{c/2})^{N-1}\psi_1^*& \cdots&-\psi_N^*[N-1,N-1] & -(i\sqrt{c/2})^{N-1}\phi_1^*   &\cdots&-\phi_1^*[N-1,N-1]\\\cdots&\cdots&\cdots&\cdots&\cdots&\cdots&\cdots&\cdots&\cdots\\\varphi_1 & \cdots& \varphi_1[0,N-1]& -\psi_1^* & \cdots &-\psi_1^*[0,N-1]& -\phi_1^* & \cdots& -\phi_1^*[0,N-1]
\end{array} \right|}\nonumber
\end{eqnarray}
\begin{eqnarray}
&&D_2 =\nonumber\\&&{\tiny\left|\begin{array}{ccccccccc}
 (i\sqrt{c/2})^{N-1}\psi_1& \cdots &\psi_1[N-1,N-1]& (i\sqrt{c/2})^{N-1} \varphi_1^*& \cdots& \varphi_1^*[N-1,N-1]& 0 &\cdots&0\\
\cdots&\cdots&\cdots&\cdots&\cdots&\cdots&\cdots&\cdots&\cdots\\\psi_1 & \cdots &\psi_1[0,N-1] & \varphi_1^* & \cdots& \varphi_1^*[0,N-1]& 0 &\cdots&0\\(i\sqrt{c/2})^{N-1}\phi_1&\cdots&\phi_1 [N-1,N-1]& 0& \cdots&0 &\varphi_1^* (i\sqrt{c/2})^{N-1}& \cdots&\varphi_1^*[N-1,N-1] \\\cdots&\cdots&\cdots&\cdots&\cdots&\cdots&\cdots&\cdots&\cdots\\\phi_1 & \cdots& \phi_1[0,N-1]& 0 &\cdots&0 & \varphi_1^* & \cdots& \varphi_1^*[0,N-1]\\
 (i\sqrt{c/2})^{N}\psi_1& \cdots &\psi_1[N,N-1]& (i\sqrt{c/2})^{N} \varphi_1^*& \cdots& \varphi_1^*[N,N-1]& 0 &\cdots&0\\\cdots&\cdots&\cdots&\cdots&\cdots&\cdots&\cdots&\cdots&\cdots\\\varphi_1 & \cdots& \varphi_1[0,N-1]& -\psi_1^* & \cdots &-\psi_1^*[0,N-1]& -\phi_1^* & \cdots& -\phi_1^*[0,N-1]
\end{array} \right|}\nonumber
\end{eqnarray}
\begin{eqnarray}
&&D_3 =\nonumber\\&&{\tiny\left|\begin{array}{ccccccccc}
 (i\sqrt{c/2})^{N-1}\psi_1& \cdots &\psi_1[N-1,N-1]& (i\sqrt{c/2})^{N-1} \varphi_1^*& \cdots& \varphi_1^*[N-1,N-1]& 0 &\cdots&0\\
\cdots&\cdots&\cdots&\cdots&\cdots&\cdots&\cdots&\cdots&\cdots\\\psi_1 & \cdots &\psi_1[0,N-1] & \varphi_1^* & \cdots& \varphi_1^*[0,N-1]& 0 &\cdots&0\\(i\sqrt{c/2})^{N-1}\phi_1&\cdots&\phi_1 [N-1,N-1]& 0& \cdots&0 &\varphi_1^* (i\sqrt{c/2})^{N-1}& \cdots&\varphi_1^*[N-1,N-1] \\\cdots&\cdots&\cdots&\cdots&\cdots&\cdots&\cdots&\cdots&\cdots\\\phi_1 & \cdots& \phi_1[0,N-1]& 0 &\cdots&0 & \varphi_1^* & \cdots& \varphi_1^*[0,N-1]\\
(i\sqrt{c/2})^{N}\phi_1&\cdots&\phi_1 [N,N-1]& 0& \cdots&0 &\varphi_1^* (i\sqrt{c/2})^{N}& \cdots&\varphi_1^*[N,N-1]\\\cdots&\cdots&\cdots&\cdots&\cdots&\cdots&\cdots&\cdots&\cdots\\\varphi_1 & \cdots& \varphi_1[0,N-1]& -\psi_1^* & \cdots &-\psi_1^*[0,N-1]& -\phi_1^* & \cdots& -\phi_1^*[0,N-1]
\end{array} \right|}\nonumber
\end{eqnarray}
where
\begin{eqnarray}
\psi_1[j,n]&=&\frac{1}{(n)!}\frac{\partial^{n}}{\partial \lambda_1^{n}}[(\lambda_1+\delta)^j\psi_1(\lambda_1+\delta)]|_{\delta=0},\nonumber\\
\phi_1[j,n]&=&\frac{1}{(n)!}\frac{\partial^{n}}{\partial \lambda_1^{n}}[(\lambda_1+\delta)^j\phi_1(\lambda_1+\delta)]|_{\delta=0},\nonumber\\
\varphi_1[j,n]&=&\frac{1}{(n)!}\frac{\partial^{n}}{\partial \lambda_1^{n}}[(\lambda_1+\delta)^j\varphi_1(\lambda_1+\delta)]|_{\delta=0}.
\label{sd1331}
\end{eqnarray}
Evaluating the determinants in (\ref{gdtnth}) we can get the N-th order RW solution of GCNLS system (\ref{pct1}) through GDT with plane wave solution as the seed solution.    
\section{Multi-RW solutions of GCNLS system}
\label{RW}
In the previous section, we have derived the necessary formula  to generate $N$-th order RW solution for the GCNLS system (\ref{pct1}).  In this section, using this description, we construct explicit multi-RW solutions of the GCNLS Eq. (\ref{pct1}).
\subsection{First order RW solution}
We begin our analysis with plane wave solutions as the seed solution, that is $p[0]=a_1e^{ic_1t}$ and $q[0]=a_2e^{ic_2t}$, where $a_1$, $a_2$ and $c_1$ and $c_2$ are real constants.  Using the above forms in Eq. (\ref{pct1}) and restricting $c_1=c_2\equiv c$ we obtain a consistent dispersion relation of the form $\frac{c}{2}=a_1^2+a_2^2+(b+b^*)a_1a_2$.  Substituting the above seed solution into the Lax pair equations (\ref{pct256}) and solving the resultant system of equations we obtain the following special solution with $\lambda_1=ih$, namely
\begin{eqnarray} 
\Psi_1(\lambda_1)=\left(\begin{array}{ccc}(k_1\frac{a_1}{h+\sqrt{h^2-c/2}}e^A+k_2\frac{a_1}{h-\sqrt{h^2-c/2}}e^{-A})e^{\frac{ic}{2}t}\\(k_1\frac{a_2}{h+\sqrt{h^2-c/2}}e^A+k_2\frac{a_2}{h-\sqrt{h^2-c/2}}e^{-A})e^{\frac{ic}{2}t}\\(k_1e^A+k_2e^{-A})e^{\frac{-ic}{2}t}\end{array}\right),
\label{bassol}
\end{eqnarray}
where $A=\sqrt{h^2-\frac{c}{2}}(x-2iht+\Phi(f))$, where $\Phi(f)=\sum_{i=0}^Ns_if^{2i},\;\; s_i\;\epsilon\; \mathbb{C}$,
\begin{eqnarray}
{\text{and}}\;\; k_1=\frac{2c[(h-\sqrt{h^2-c/2})^2-c/2]}{(c-2h^2)(h-\sqrt{h^2-c/2})}, \;\; k_2=\frac{2c[(h+\sqrt{h^2-c/2})^2-c/2]}{(c-2h^2)(h+\sqrt{h^2-c/2})}.
\label{bassol1}
\end{eqnarray} 
We have also included an arbitrary parameter $\Phi(f)$ in the phase factor to obtain  triplet, triangular and hexagonal structures of RWs.  Plugging the above basic solutions, (\ref{bassol}) and (\ref{bassol1}), in the first iterated DT formula (\ref{pct9}) we can obtain the Akhmediev breather solution.  
\par To obtain the RW solution we fix the critical eigenvalue to be, $\lambda_1=\lambda_0=ih=i(\sqrt{c/2}+f^2)$. 
We expand the critical eigenfunction $\Psi_1(f)$ at $f=0$ (vide Eq. (\ref{g1})) to obtain
\begin{eqnarray}
\Psi_2=\Psi_1(\lambda_1+f^2)=\Psi_1^{[0]}+\Psi_1^{[1]}f^2+\Psi_1^{[2]}f^4+...+\Psi_1^{[N]}f^{2N}+...,
\label{psi2}
\end{eqnarray}
\begin{figure}[H]
\includegraphics[width=1\linewidth]{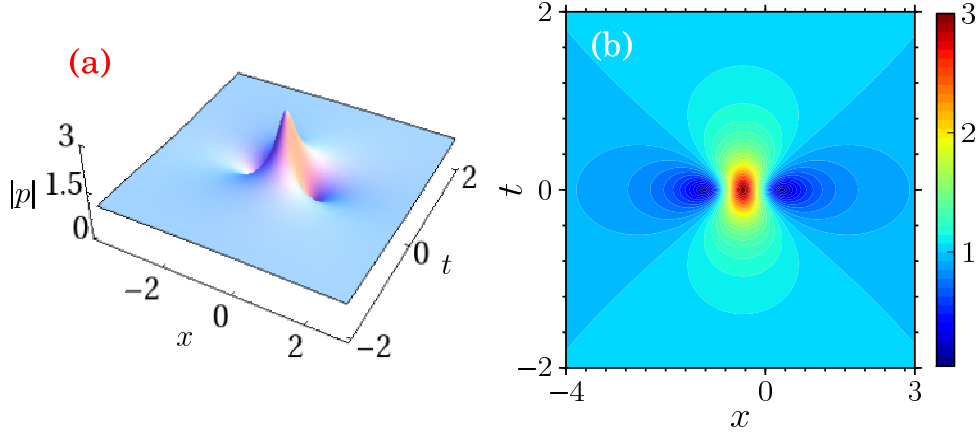}
\caption{(a) First order RW solution of the $p$ component, (b) Corresponding contour plot.  Similar profile occurs for $q$ also (not shown here).}
\end{figure}
where
\begin{eqnarray}
\Psi_1^{[0]}&=&\left(\begin{array}{ccc}8a_1(-i\sqrt{2c}t+x)e^{\frac{ic}{2}t}\\8a_2(-i\sqrt{2c}t+x)e^{\frac{ic}{2}t}\\(8-8ict+4\sqrt{2c}x)e^{\frac{-ic}{2}t}\end{array}\right),\nonumber\\
\Psi_1^{[1]}&=&\left(\begin{array}{ccc}\frac{8}{3}a_1[6l+6im-12it+4ic^2t^3-6\sqrt{2}c^{3/2}t^2x-6ictx^2+\sqrt{2c}x^3]e^{\frac{ic}{2}t}\\\frac{8}{3}a_2[6l+6im-12it+4ic^2t^3-6\sqrt{2}c^{3/2}t^2x-6ictx^2+\sqrt{2c}x^3]e^{\frac{ic}{2}t}\\\frac{8}{3}[3\sqrt{2c}l+3i\sqrt{2c}m-12i\sqrt{2c}t-6\sqrt{2}c^{3/2}t^2+(6-12ict-6c^2t^2)x\nonumber\\+2i\sqrt{2}c^{5/2}t^3+3\sqrt{2c}x^2-3i\sqrt{2}c^{3/2}tx^2+cx^3]e^{\frac{-ic}{2}t}\end{array}\right),\nonumber\\
\Psi_1^{[2]}&=&\left(\begin{array}{ccc}&&\frac{16}{5}e^{\frac{ict}{2}}(\Psi_1^{[21]})\\
&&\frac{16}{5}e^{\frac{ict}{2}}(\Psi_1^{[22]})\\
&&\frac{8}{5}e^{-\frac{ict}{2}}(\Psi_1^{[23]})\end{array}\right),...
\label{psi10}
\end{eqnarray}
with
\begin{eqnarray}
\Psi_1^{[21]}&=&a_1(60e+60ig-60\sqrt{2}c^{3/2}lt^2-60i\sqrt{2}c^{3/2}mt^2+140i\sqrt{2}c^{3/2}t^3-4i\sqrt{2}c^{7/2}t^5
\nonumber\\&&+120cmtx-300ct^2x+20c^3t^4x+30\sqrt{2c}lx^2+30i\sqrt{2c}mx^2-90i\sqrt{2c}tx^2\nonumber\\&&+20i\sqrt{2}c^{5/2}t^3x^2+10x^3-20c^2t^2x^3-5i\sqrt{2}c^{3/2}tx^4+cx^5-120icltx),\nonumber\\
\Psi_1^{[22]}&=&a_2(60e+60ig-60\sqrt{2}c^{3/2}lt^2-60i\sqrt{2}c^{3/2}mt^2+140i\sqrt{2}c^{3/2}t^3-4i\sqrt{2}c^{7/2}t^5
\nonumber\\&&+120cmtx-300ct^2x+20c^3t^4x+30\sqrt{2c}lx^2+30i\sqrt{2c}mx^2-90i\sqrt{2c}tx^2\nonumber\\&&+20i\sqrt{2}c^{5/2}t^3x^2+10x^3-20c^2t^2x^3-5i\sqrt{2}c^{3/2}tx^4+cx^5-120icltx),\nonumber
\end{eqnarray}
\begin{eqnarray}
\Psi_1^{[23]}&=&60\sqrt{2c}e+60i\sqrt{2c}g+120(l+im)-240it-240iclt+240cmt-600ct^2\nonumber\\&&
-120ic^2mt^2+360ic^2t^3+40c^3t^4-8ic^4t^5+120\sqrt{2c}lx+120i\sqrt{2c}mx\nonumber\\&&
-360i\sqrt{2c}tx-120i\sqrt{2}c^{3/2}ltx+120\sqrt{2}c^{3/2}mtx-420\sqrt{2}c^{3/2}t^2x\nonumber\\&&
+80i\sqrt{2}c^{5/2}t^3x+20\sqrt{2}c^{7/2}t^4x+60x^2+60clx^2+60icmx^2-300ictx^2\nonumber\\&&-120c^2t^2x^2+40ic^3t^3x^2-120c^2lt^2
+30\sqrt{2c}x^3-40i\sqrt{2}c^{3/2}tx^3-20\sqrt{2}c^{5/2}t^2x^3\nonumber\\&&+10cx^4+\sqrt{2}c^{3/2}x^5.
\label{psi121}
\end{eqnarray}
\par Substituting $\Psi_1^{[0]}$ in (\ref{g2}), we can obtain the first order RW solution in the form 
\begin{eqnarray}
p[1]&=&\frac{a_1e^{ict}(-6c^2t^2-2\sqrt{2c}x+c(8it+4(a_1^2+a_2^2)t^2-3x^2)+2(1+a_1^2x^2+a_2^2x^2))}{2+2c^2t^2+2\sqrt{2c}x+cx^2+2(a_1^2+a_2^2)(2ct^2+x^2)},\nonumber\\
q[1]&=&\frac{a_2e^{ict}(-6c^2t^2-2\sqrt{2c}x+c(8it+4(a_1^2+a_2^2)t^2-3x^2)+2(1+a_1^2x^2+a_2^2x^2))}{2+2c^2t^2+2\sqrt{2c}x+cx^2+2(a_1^2+a_2^2)(2ct^2+x^2)}.\nonumber\\
\end{eqnarray}
One may note that the components $p$ and $q$ are proportional to each other.  The first order RW solution is plotted in Fig.1, which is localized in both $x$ and $t$.
\subsection{Second order RW solution}
\begin{figure}[H]
\includegraphics[width=1\linewidth]{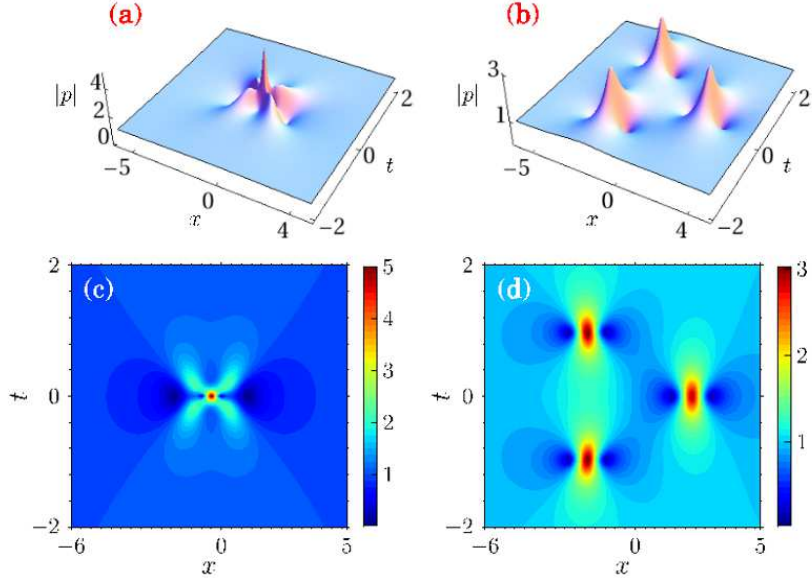}
\caption{(a) Second order RW solution of the $p$ component for the values $l=0$ and $m=0$, (c) Corresponding contour plot.  (b) Second order RW solution of the $p$ component for the values $l=15$ and $m=0$, (d) Corresponding contour plot.  Similar profile occurs for $q$ also (not shown here).}
\end{figure} 
\begin{figure}[H]
\includegraphics[width=1\linewidth]{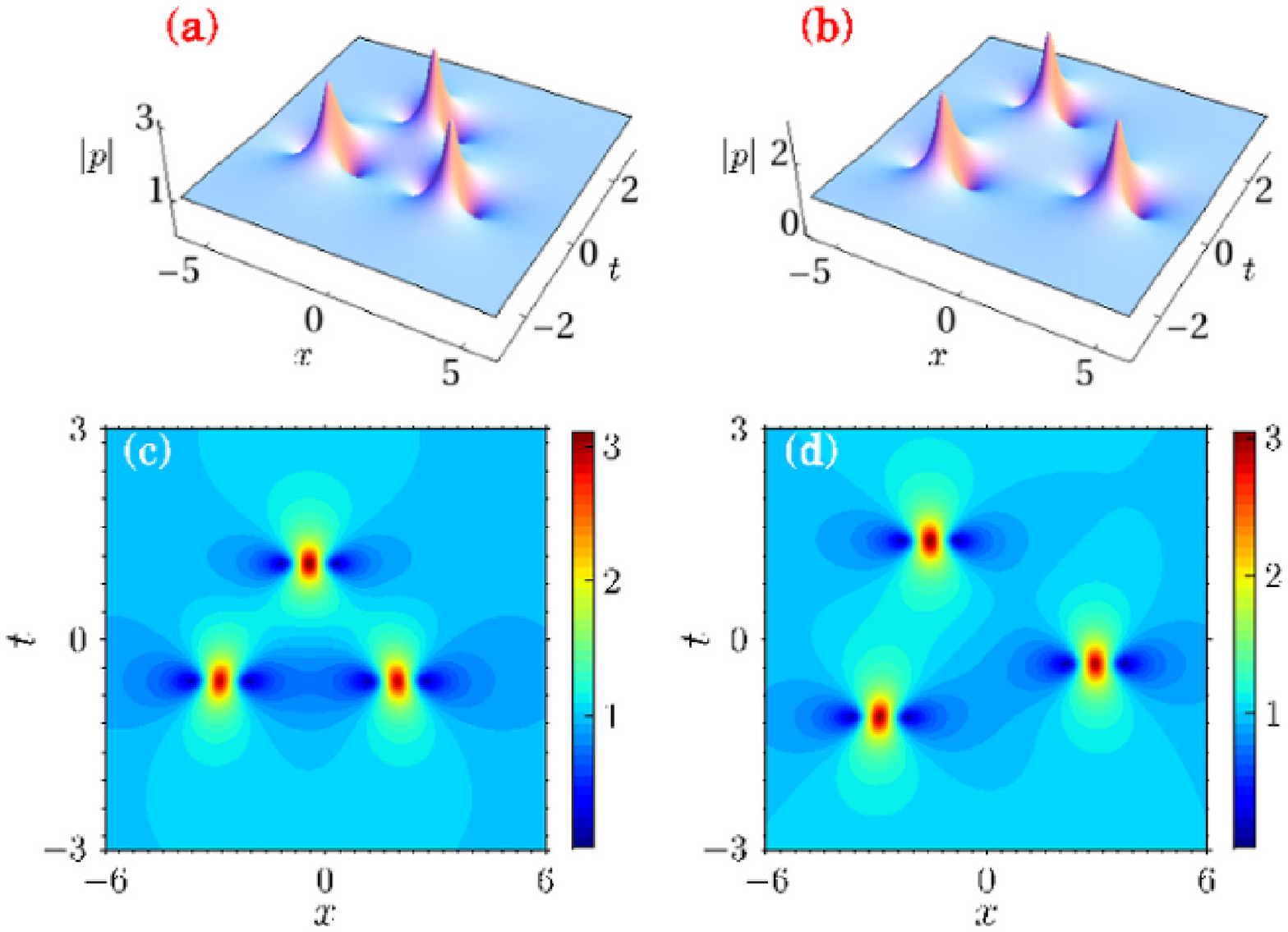}
\caption{(a) Second order RW solution of the $p$ component for the values $l=0$ and $m=15$, (c) Corresponding contour plot.  (b) Second order RW solution of the $p$ component for the values $l=15$ and $m=15$, (c) Corresponding contour plot.  Similar profile occurs for $q$ also (not shown here).}
\end{figure} 
\par To obtain the second order RW solution, we use the limit approach given in Eq. (\ref{g3}), that is 
\begin{eqnarray}
\Psi_2[1]&=&{\lim_{f \rightarrow 0}}\frac{[if^2+T_1[1]]\psi_1(f)}{f^2}\nonumber\\
&=&T_1[1]\Psi_1^{[1]}+i\Psi_1[0]\equiv \Psi_1[1]
\label{psi11}
\end{eqnarray}
\begin{eqnarray}
{\text {with}}\;\; T_1[1]=\lambda_1I-H_1[0]\Lambda[1] H_1[0]^{-1}.
\end{eqnarray}
Since we  know the expressions $\Psi_1^{[0]}$ and $\Psi_1^{[1]}$ through (\ref{psi2})-(\ref{psi121}) and $T_1[1]$ (vide Eq. (\ref{pct5})) as well, we can get the exact form of $\Psi_1[1]=(\psi_1[1],\phi_1[1],\varphi_1[1])^T$ readily.  Substituting $\Psi_1[1]$ in Eq. (\ref{2ndgdt}) we can obtain the second order RW solution.  On the other hand we can also get the second order RW solution from the determinant expressions itself (vide Eq. (\ref{sd133})).  In both the ways we end up at
\begin{eqnarray}
p[2]=12a_1e^{2i(a_1^2+a2^2)t}\;\frac{\sum_{n=0}^{6}N_{nr}x^n+i\sum_{n=0}^{6}N_{ni}x^n}{D_2},\nonumber\\
q[2]=12a_2e^{2i(a_1^2+a2^2)t}\;\frac{\sum_{n=0}^{6}N_{nr}x^n+i\sum_{n=0}^{6}N_{ni}x^n}{D_2}.
\label{2ndsol}
\end{eqnarray}
We have given the exact expressions of $N_{nr}$, $N_{ni}$ and $D_2$ in Appendix A.
 
\par The results are shown in Figs. 2 and 3.  The second order RW solution is derived with two free parameters, namely $l$ and $m$.  We analyze the second order RW solution based on these two free parameters.  When $l=m=0$, we have the classical second order RW solution which is demonstrated in Figs. 2(a) and 2(c).  It contains one largest crest and four subcrests with two deepest troughs.  The RW gets deformed when we increase the values to $l=.3$, $m=0$ or $l=0$, $m=0.3$ or $l=m=0.2$.  Increasing the parameter values further we observe that the second order RW splits into three first order RWs and they emerge in a triangular form which is known to be a triplet pattern.  We observe the formation of triplet structure at $l=3$, $m=0$ and $l=0$, $m=2.8$.  When we increase the free parameter values further the distance between the peaks in the triplet increases.  Figs. 2(b) and 2(d) display the triplet plot of second order RW for the value $l=15$ and $m=0$.  When we interchange the values of $l$ and $m$, say for example $l=0$ and $m=15$, the triangular pattern still persists but the three peaks now appear in a different orientation which is demonstrated in Figs. 3(a) and 3(c).  Finally, when we increase the values of both the parameters $l$ and $m$ to $10$, we get the same triangular pattern but in a different orientation, which is displayed in Figs. 3(b) and 3(d). 

\subsection{Third order RW solution}
\par We proceed to construct the third order RW solution of GCNLS system (\ref{pct1}) through the limit process (\ref{pctpsi12}).  Doing so, we find 
\begin{eqnarray}
\Psi_1[2]=-\Psi_1^{[0]}+i[T_1[1]+T_1[2]]\Psi_1^{[1]}+T_1[2]T_1[1]\Psi_1^{[2]}.
\label{pctpsi121}
\end{eqnarray}
Substituting the data given in (\ref{psi2})-(\ref{psi121}) into (\ref{pctpsi12}) we can obtain the explicit expression of $\Psi_1[2]$.  Plugging the latter into the solution (\ref{pct3rd}) we arrive at the third order RW solution for the GCNLS system (\ref{pct1}).  Since the explicit expression of third order RW solution is very lengthy we are not presenting the obtained form here.  However, in the following, we analyze the third order RW solution graphically.
\begin{figure}[H]
\includegraphics[width=1\linewidth]{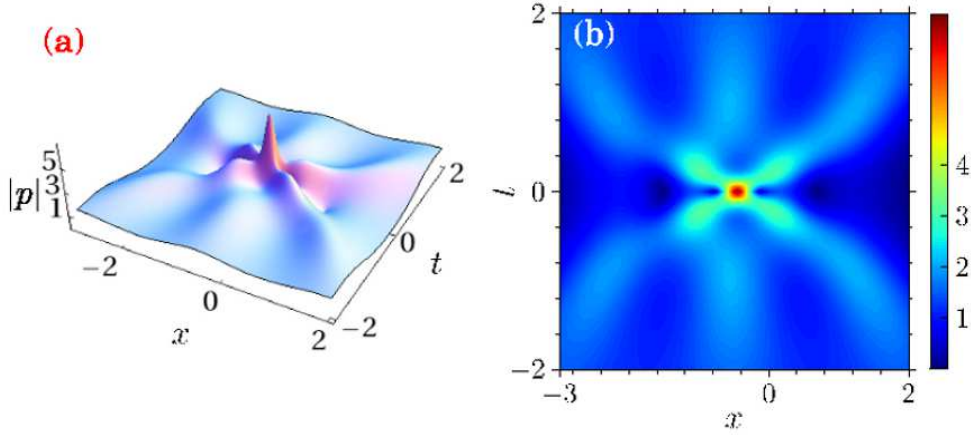}
\caption{(a)Third order RW solution of the $p$ component for the values $l=0$, $m=0$, $e=0$ and $m=0$, (b) Corresponding contour plot.  Similar profile occurs for $q$ also (not shown here).}
\end{figure} 
\begin{figure}[H]
\includegraphics[width=1\linewidth]{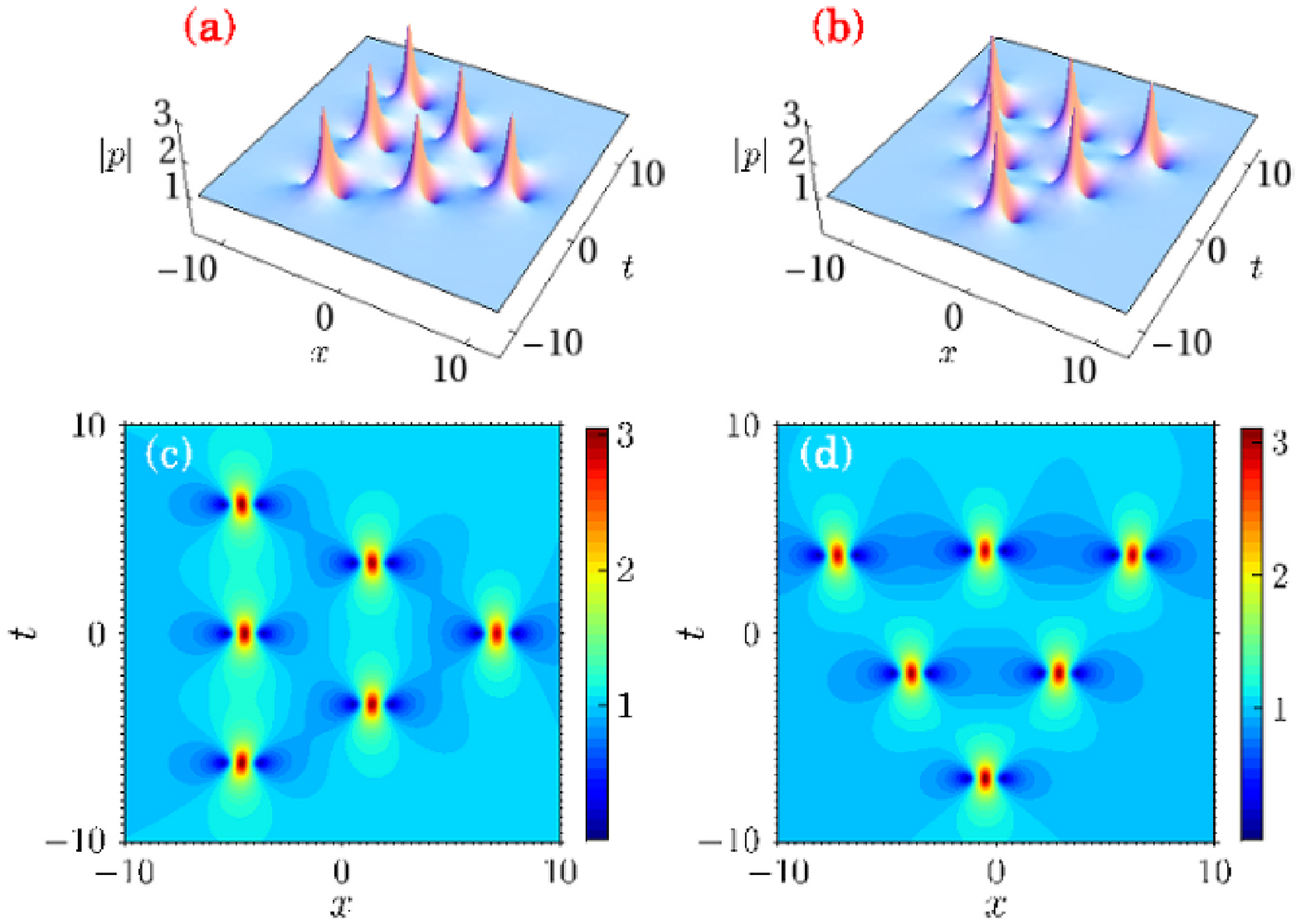}
\caption{(a) Third order RW solution of the $p$ component for the values $l=60$, $m=0$, $e=0$ and $g=0$, (c) Corresponding contour plot. (b) Third order RW solution of the $p$ component for the values  $l=0$, $m=70$, $e=0$ and $g=0$, (d) Corresponding contour plot.  Similar profile occurs for $q$ also (not shown here).}
\includegraphics[width=1\linewidth]{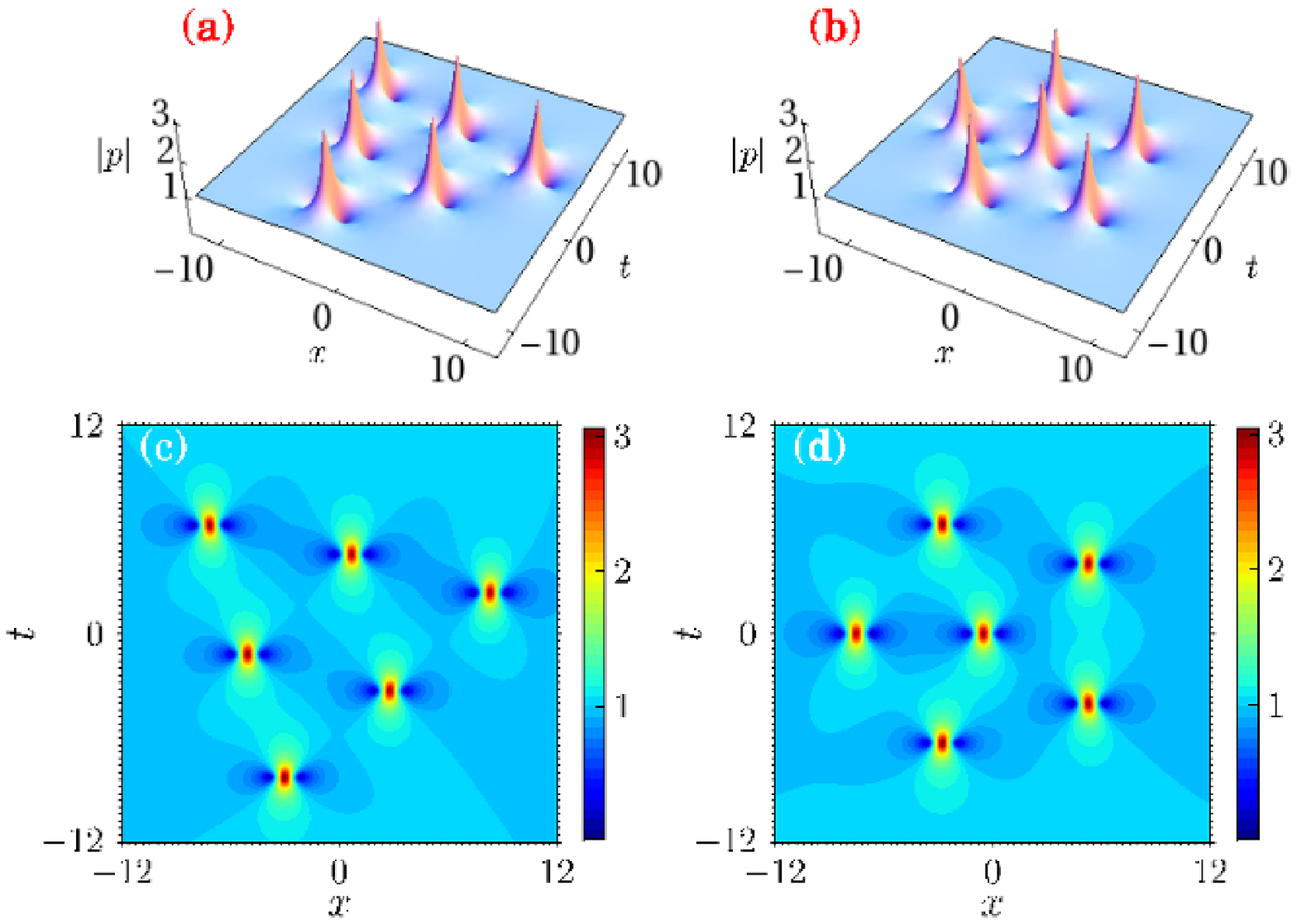}
\caption{(a) Third order RW solution of the $p$ component for the values  $l=60$, $m=70$, $e=0$ and $g=0$, (c) Corresponding contour plot.  (b) Third order RW solution of the $p$ component for the values $l=0$, $m=0$, $e=2000$ and $g=0$, (d) Corresponding contour plot.  Similar profile occurs for $q$ also (not shown here).}
\end{figure}
\begin{figure}[H]
\includegraphics[width=1\linewidth]{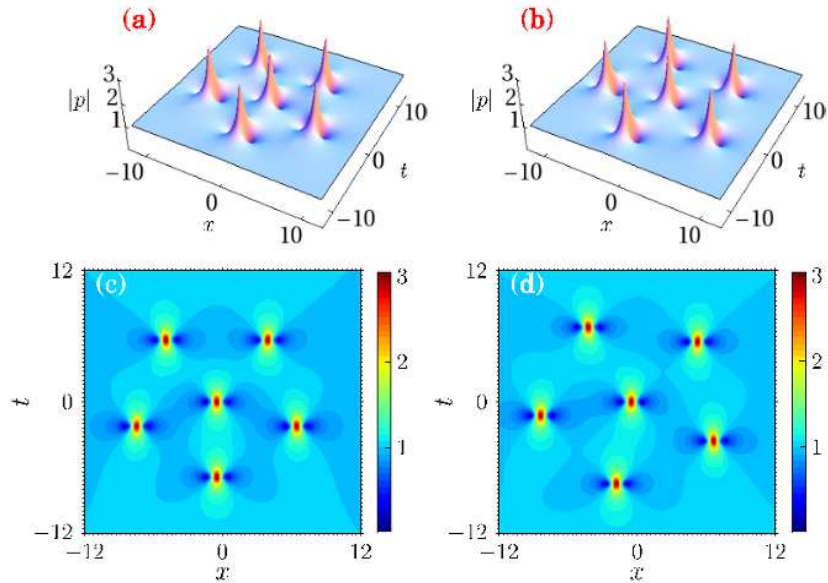}
\caption{(a) Third order RW solution of the $p$ component for the values $l=0$, $m=0$, $e=0$ and $g=2500$, (c) Corresponding contour plot.  (b) Third order RW solution of the $p$ component for the values $l=0$, $m=0$, $e=2000$ and $g=1500$, (c) Corresponding contour plot.  Similar profile occurs for $q$ also (not shown here).}
\end{figure}

\par The third order RW solution is derived with four free parameters, namely $l$, $m$, $e$ and $g$.  We analyze the formation of RW patterns with respect to these four free parameters.  To begin with, we reproduce the classical third order RW form by restricting all the free parameters are to be zero.  The resultant outcome is shown in Fig. 4.  When we increase the values of these parameters the third order RW splits into six first order RWs.  
The third order RW gets deformed in the vicinity of the following parametric choices, namely (i) $l=0.1$  $m,e,g=0$, (ii) $l,e,g=0$, $m=0.5$, (iii) $l,m,g=0$, $e=0.2$, (iv) $l,m,e=0$, $g=0.5$ and (v) $l,m,e,g=0.1$.  For large values of $l$ and $m$ and small values of $e$ and $g$ we observe triangular structure with six peaks.  On the other hand for large values of $e$ and $g$ and small values of $l$ and $m$ we get a ring structure with same number of peaks.  In Figs. 5(a) and 5(c), we display the third order RW solution for $l=60$, $m,e,g=0$, in which we observe a triangular pattern with six first order RWs.  In  Figs. 5(b) and 5(d) we depict the same solution for $l,e,g=0$ $m=70$, in which we observe the same triangular pattern but the six peaks now appear in a different orientation.  We increase the values to $l=60$, $m=70$, $e=g=0$ and display the outcome in Figs. 6(a) and 6(c).  The triangular structure still persists but the peaks assemble in a different orientation.

When we investigate the structure of third order RW  solution with $l,m,g=0$, $e=2000$, we obtain a ring structure with six peaks as shown in Figs. 6(b) and 6(d).  On the other hand when we interchange the values of $e$ and $g$ as $e=0$ and $g=2500$ with $l=m=0$ we again have the ring pattern but in a different orientation which is shown in Figs. 7(a) and 7(c).  Finally, we plot the solution for $l=m=0$, $e=2000$, $g=1500$ and present the outcome in Figs. 7(b) and 7(d) in which the ring structure is observed similar to Figs. 7(a) and 7(c) but in a different orientation.  From these plots, we infer that even for small variations in $l$ and $m$ (with $e=g=0$) we can observe the triangular pattern.  On the other hand the ring pattern can be visualized only for large values of $e$ and $g$ with $l=m =0$.   

\section{Conclusion}
\label{conclusion}
In this paper we have discussed the method of constructing N-th order RW solution for the GCNLS system (\ref{pct1}).  Since it is very difficult to construct N-th order RW solution through conventional DT we have adopted the GDT method and presented a recursive formula for the N-th order RW solution.  We have given the explicit form of first and second order RW solutions.  Since the third order RW solution is very lengthy we have given only the solution formula and determinant representation of it.  However, we have analyzed the third order RW profile graphically in detail.  We have derived the second order RW solution with two free parameters and the third order RW solution with four free parameters respectively.  These solutions satisfy the original equation when the four wave mixing coefficient becomes pure imaginary.  We have also analyzed the second and third order RW solutions by varying these free parameters and obtained certain interesting structures exhibited by them.  For example, in the case of second order RW, we have shown that these RWs exhibit a triplet pattern.  As far as the third order RW solution is concerned we have four free parameters, namely  $l,m,e$ and $g$.  We have captured the classical RW solution when all these free parameters are zero.  We have visualized a  triangular pattern for certain non-zero values of $l$ and $m$ with $e,g=0$.  We have also observed that these RWs exhibit a hexagonal structure for $l,m=0$ and $e,g\neq 0$.  In addition to the above, we have given the determinant representation of N-th order RW solution which will be useful to generate higher order RWs through symbolic manipulation program.  The N-th order RW solution contains $2N-2$ free parameters which will be useful again to generate certain interesting patterns that persist in higher order RWs.  The results obtained in this paper will be useful in the study of rogue waves in birefringent optical fibers, multi-component Bose-Einstein condensates, multi-component plasmas and so on.   

\section*{Acknowledgements}                      
NVP wishes to thank the University Grants Commission (UGC-RFSMS), Government of India, for providing a Research Fellowship. The work of MS forms part of a research project sponsored by National Board for Higher Mathematics (NBHM), Government of India.
\appendix
\section{Forms of $N_{nr}$, $N_{ni}$ and $D_2$ of second order RW solution}
In the following we provide the exact expressions of $N_{nr}$, $N_{ni}$, $n=0,1,\cdots 6$ and which appear in (\ref{2ndsol}).
\begin{eqnarray}
N_{0r}&=&3Al-24A^2mt-12A^2(-8+5Al)t^2+144A^4mt^3-160A^4t^4-1152A^6t^6\nonumber\\
N_{1r}&=&3\sqrt{A}(l+4Al)-60A^{5/2}mt-12A^{5/2}(-19+14Al)t^2+96A^{9/2}mt^3\nonumber\\&&-256A^{9/2}t^4-512A^{13/2}t^6\nonumber\\
N_{2r}&=&3A(4+5Al)-60A^3mt-24A^3(-9+4Al)t^2\nonumber\\
N_{3r}&=&A^{3/2}(19+6Al)-24A^{7/2}mt+64A^{7/2}-256A^{11/2}t^4\nonumber\\
N_{4r}&=&16A^2-56A^4t^2\nonumber\\
N_{5r}&=&8A^{5/2}-32A^{9/2}t^2\nonumber\\
N_{6r}&=&2A^3\nonumber\\
N_{0i}&=&3Am+12A(-1+2Al)t-84A^3mt^2-16A^3(-17+3Al)t^3+96A^5mt^4\nonumber\\&&+640A^5t^5-512A^7t^7\nonumber\\
N_{1i}&=&12A^{3/2}m+12A^{3/2}(-2+7Al)t-120A^{7/2}mt^2-16A^{7/2}(-35+6Al)t^3\nonumber\\&&+640A^{11/2}t^5\nonumber\\
N_{2i}&=&21A^2m+84A^3lt-48A^4mt^2+608A^4t^3-128A^6t^5\nonumber\\
N_{3i}&=&18A^{5/2}m+4A^{5/2}(11+6Al)t+320A^{9/2}t^3\nonumber\\
N_{4i}&=&6A^3m+64A^3t+32A^5t^3\nonumber\\
N_{5i}&=&40A^{7/2}t\nonumber\\
N_{6i}&=&8A^4t
\end{eqnarray}
\begin{eqnarray}
D_2&=&\left(\sum_{n=0}^{7}D_{nr}x^n+i\sum_{n=0}^{8}D_{ni}x^n\right)
\end{eqnarray}
\begin{eqnarray}
D_{0r}&=&36A(1+4A^2(l^2+m^2))t-576A^3mt^2+1152A^3(1+Al)t^3\nonumber\\&&-1536A^5mt^4+7680A^5t^5+4096A^7t^7\nonumber\\
D_{1r}&=&36A^{3/2}(5+4A^2(l^2+m^2))t+576A^{7/2}mt^2+3456A^{9/2}lt^3\nonumber\\&&-1536A^{11/2}mt^4+10752A^{11/2}t^5+4096A^{15/2}t^7\nonumber\\
D_{2r}&=&-144A^2(-3+2Al)t+2304A^4mt^2+1152A^4(-1+2Al)t^3\nonumber\\&&+6144A^6t^5\nonumber\\
D_{3r}&=&-96A^{5/2}(-6+5Al)t+1152A^{9/2}mt^2+1536A^{9/2}t^3+3072A^{13/2}\nonumber\\
D_{4r}&=&-192A^3(-3+Al)t+2304A^5t^3\nonumber\\
D_{5r}&=&96A^{7/2}(5+8A^2t^2)t\nonumber\\
D_{6r}&=&256A^4t\nonumber\\
D_{7r}&=&64A^{9/2}t\nonumber\\
\end{eqnarray}
\begin{eqnarray}
D_{0i}&=&-9(1+4A^2(l^2+m^2))+36A^2(-7+4A(1(-2+Al)+Am^2))t^2\nonumber\\&&-192A^4mt^3+384A^4(-2+3Al)t^4-1536A^6mt^5+6656A^6t^6+4096A^8t^8\nonumber\\
D_{1i}&=&-18\sqrt{A}(3+4A^2(l^2+m^2))-144A^{5/2}(1+8Al)t^2+1920A^{9/2}A^4mt^3\nonumber\\&&+1152A^{9/2}(-5+2Al)t^4+1024A^{3/2}t^6\nonumber\\
D_{2i}&=&-9A(17+4A(l(-2+Al)+Am^2))-720A^3mt-576A^4(-1+3Al)t^2\nonumber\\&&+1536A^5mt^3-4224A^5t^4+2048A^7t^6\nonumber\\
D_{3i}&=&12A^{3/2}(-21+16Al)-864A^{7/2}mt-192A^{7/2}(-1+4Al)t^2-768A^{11/2}t^4\nonumber\\
D_{4i}&=&24A^2(-12+7Al)-288A^4mt-672A^4t^2\nonumber\\
D_{5i}&=&24A^{5/2}(-11+2Al)-576A^{9/2}t^2\nonumber\\
D_{6i}&=&-184A^3-128A^5t^2\nonumber\\
D_{7i}&=&-80A^{7/2}\nonumber\\
D_{8i}&=&-16A^4
\end{eqnarray}
\appendix

\end{document}